\documentclass[superscriptaddress,groupedaddress,nofootnoteinbib,12pt]{article}  

\usepackage{graphicx}  
\usepackage{dcolumn}   
\usepackage{bm}        
\usepackage{jcappub}

\def\be{\begin{equation}}
\def\ee{\end{equation}}
\def\ba{\begin{eqnarray}}
\def\ea{\end{eqnarray}}
\def\beq{\begin{eqnarray}}
\def\eeq{\end{eqnarray}}

\def\mpl{M_{\rm Pl}}

\def\E{\mathcal{E}}

\def\d{\mathrm{d}}
\def\p{{\cal P}}

\def\L*{{\cal L}_*}
\def\L{\mathcal{L}}
\def\({\left(}
\def\){\right)}

\def\nn{\nonumber}
\def\p{\partial}

\def\stu{St\"uckelberg }
\def\poin{Poincar\'{e} }
\def\p{\partial}

\def\<{\langle}
\def\>{\rangle}

\def\pa {\partial}

\def\cs2{c_{s}^{2}}

 \def\al{\alpha}

 \def\de{\delta}
 
 \def\ep{\varepsilon}

 \def\la{\lambda}
 \def\La{\Lambda}
 \def\si{\sigma}

 \def\om{\omega}
 \def\Om{\Omega}
 
 \def\p{\partial}

 \def\wed{\wedge}

\def\Mpl{M_{\rm Pl}}
\def\mpl{M_{\rm Pl}}

 \def\be   {\begin{equation}}   \def\ee   {\end{equation}}

 \def\ba  {\begin{eqnarray}}   \def\ea  {\end{eqnarray}}

\begin{document}

\title{Interactions of Charged Spin-2 Fields}

\author{Claudia de Rham, Andrew Matas, Nicholas Ondo}
\author{and Andrew J.~Tolley}
\affiliation{CERCA/Department of Physics, Case Western Reserve University, 10900 Euclid Ave, Cleveland, OH 44106, USA}


\abstract{In light of recent progress in ghost-free theories of massive gravity and multi-gravity, we reconsider the problem of constructing a ghost-free theory of an interacting spin-2 field charged under a $U(1)$ gauge symmetry. Our starting point is the theory originally proposed by Federbush, which is essentially Fierz-Pauli generalized to include a minimal coupling to a $U(1)$ gauge field. We show the Federbush theory with a dynamical $U(1)$ field is in fact ghost-free and can be treated as a healthy effective field theory to describe a massive charged spin-2 particle. It can even potentially have healthy dynamics above its strong-coupling scale. We then construct candidate gravitational extensions to the Federbush theory both by using Dimensional Deconstruction, and by constructing a general non-linear completion. However, we find that the $U(1)$ symmetry forces us to modify the form of the Einstein-Hilbert kinetic term. By performing a constraint analysis directly in the first-order form, we show that these modified kinetic terms inevitably reintroduce the Boulware-Deser ghost. As a by-product of our analysis, we present a new proof for ghost-freedom of bi-gravity in 2+1 dimensions (also known as Zwei-Dreibein gravity). We also give a complementary algebraic argument that the Einstein-Hilbert kinetic term is incompatible with a $U(1)$ symmetry, for a finite number of gravitons.}

\maketitle

\section{Introduction}

It has been more than seventy years since Wigner demonstrated that all consistent, relativistic, quantum particles can be classified by their mass $m$ and their spin $j$ \cite{Wigner:1939cj, Weinberg:1995mt}. Experimentally, particle accelerators have established the existence of composite, charged massive higher spin particles \cite{Beringer:1900zz}. Nevertheless, the theoretical understanding of higher spin fields is considerably less developed than their lower spin counterparts.

The most obvious bosonic higher spin theory to consider is spin-2.
There are
arguments that the only consistent theory of a massless, self-interacting, Lorentz-invariant spin-2 field
is General Relativity \cite{Gupta:1954zz,Weinberg:1965rz,Deser:1969wk,Feynman:1996kb,Boulware:1974sr}. In fact, recent work has established that these assumptions may be weakened somewhat. Ghost-freedom alone is sufficient to derive the Einstein-Hilbert action as the kinetic term for Lorentz-invariant massive fields \cite{deRham:2013tfa} or for massless gravity theories where time translation invariance is broken explicitly \cite{Khoury:2013oqa,Khoury:2014sea}.

However the massive case is less well understood. In the 1930's, Fierz and Pauli wrote down the linearized, non-interacting theory of a single massive spin-$2$ field \cite{Fierz:1939ix, Fierz:1939zz}. There are several issues. The first is the vDVZ discontinuity of this model \cite{vanDam:1970vg, Zakharov:1970cc}.  The vDVZ discontinuity is a curious feature of the Fierz-Pauli action, which is that in the limit $m \to 0$, the Fierz-Pauli predictions do not become equivalent to that of the linearized Einstein-Hilbert Lagrangian. Vainshtein was the first to see that this discontinuity could be avoid by adding self-interactions for the massive spin-$2$ field, and associating the regime of validity of the linear approximation \cite{Vainshtein:1972sx}. However, Boulware and Deser showed that generically a non-linear extension of the Fierz-Pauli action would introduce a sixth ghost mode, the Boulware-Deser ghost \cite{Boulware:1973my}.

Only recently has a theory of a Lorentz-invariant self-interacting, massive spin-$2$ field that propagates $2(2)+1 = 5$ healthy degrees of freedom (dofs) been found \cite{deRham:2010gu, deRham:2010ik, deRham:2010kj,Hassan:2011hr, deRham:2011rn, deRham:2011qq}.  This was generalized to an arbitrary number of interacting spin-2 fields in \cite{Hassan:2011zd,Hinterbichler:2012cn}. Typically in these theories one has in mind that the graviton itself has a mass.
The theory has been applied in cosmology, where the mass of the graviton may be relevant for explaining the observed acceleration of the universe if the mass corresponds to the Hubble scale today, $m \sim H_0$, while remaining technically natural \cite{deRham:2012ew,deRham:2013qqa}. For a recent review of this types of theories, see \cite{deRham:2014zqa}.
\\

However we do not necessarily need to identify a massive spin-2 field with gravity. In this context it is interesting to think about possible additional interactions for a massive spin-2 field. A natural extension is to allow for the massive spin-2 dof to be charged under a local $U(1)$ symmetry. For example, we might try to minimally couple the spin-2 field by taking the Fierz-Pauli action for a spin-2 field $H_{\mu\nu}$ and promoting $H_{\mu\nu}$ to a complex field. We can then minimally couple $H_{\mu\nu}$ to a $U(1)$ gauge field $A_\mu$ by replacing
\be
\partial_\mu H_{\pm,\nu\sigma} \rightarrow D_\mu H_{\pm,\nu\sigma} = (\partial_\mu \mp i q A_\mu) H_{\pm \nu\sigma}\,,
\ee
where $q$ is the charge of the spin-2 field.
 We might ask if there is a consistent effective field theory description for these dofs, and whether there are consistent gravitational interactions of the massive charged spin-2 field $H_{\pm,\mu\nu}$.

In fact the minimally coupled theory of a charged spin-2 field at the linear level was studied originally by Federbush \cite{Federbush:1961}. There it was argued that there was a unique minimally coupled theory at the linear level that propagated the correct number of dofs in the background of a constant electromagnetic field.

However, Velo and Zwanziger showed that generically the minimal coupling procedure would typically lead to the presence of superluminal group and phase velocities around certain backgrounds \cite{Velo:1969bt,Velo:1970ur,Velo:1972rt}. This result was also confirmed more recently in \cite{Deser:2001dt, Porrati:2008gv}. However in light of the fact that superluminal phase and group velocities have been observed in nature, see for example \cite{2004PhRvL..93t3902B}, we should not be so quick to use this result to imply a failure of causality. Acausality only occurs if the front velocity is superluminal. At tree level this is equivalent to the velocity obtained through a characteristic analysis, but at the quantum level the computation of this velocity is strongly sensitive to the strong-coupling physics and the classical characteristic analysis cannot be trusted. A well known case of the quantum effects rendering the front velocity luminal when the low energy phase/group velocity is superluminal are the case of the propagation of light in gravitational fields \cite{Hollowood:2007kt,Shore:2007um,Hollowood:2007ku,Hollowood:2008kq}. It can be show that the effective theory obtained from integrating out the electron gives rise to superluminal phase velocities at low energies, whereas the complete one-loop photon propagator is causal.

There are several models of interacting charged spin-$2$ fields which are known to be consistent.  Recently, there has been the development of Vasiliev's higher spin theory \cite{Vasiliev:1995dn}, but an older model is that of the spin-2 resonances coming from the Kaluza-Klein tower \cite{Dolan:1983aa}.
There has also been a lot of work in constructing theories of spin-$2$ fields arising from string theory, for example, \cite{Argyres:1989cu, Porrati:2010hm, Nappi:1989ny}.
The drawback in these approaches is that they entail an infinite tower of charged, massive spin-2 fields or an infinite tower of higher spin fields.  While these are all excellent examples of UV complete theories which contain charged spin-2 states, the infinite tower structure is not palatable if we only wish to describe single meson resonances through an effective field theory. 

Charged spin-2 fields were also studied by Porrati in \cite{Porrati:2008gv,Porrati:2008an,Porrati:2008ha}. In theories of a single massive charged spin-2 field with charge $q$ coupled to a $U(1)$ gauge field there is a model-independent strong-coupling scale\footnote{Following \cite{deRham:2014wfa} we will carefully distinguish between the \emph{strong-coupling scale}, the energy at which perturbation theory breaks down, and the \emph{cutoff}, which is the mass of the lightest dof we have not included in our effective theory needed to restore unitarity. In other words we assume that some self-unitarization mechanism kicks in between the strong-coupling scale and the cutoff.} $\La_{q,3}=q^{-1/3} m$.  In other words, {\it perturbative} unitarity always breaks down at the scale $\Lambda_{q,3}$ or lower.

However the breakdown of perturbative unitarity at some scale may not require the introduction of new physics at that scale, for explicit examples see \cite{Aydemir:2012nz}. Indeed, it has been argued that the Vainshtein mechanism can act as a way of recovering unitarity non-perturbatively \cite{deRham:2014wfa}. Thus it is not necessarily appropriate to think of $\Lambda_{q,3}$ as the cutoff (meaning a scale at which new physics enters), but rather as an energy scale at which {\it perturbation theory} breaks down, i.e. the strong-coupling scale. The idea that a theory can self-unitarize above its strong-coupling scale is also the essence of the `classicalization' picture \cite{Dvali:2010jz,Dvali:2010ns}.

In this work we shall extend the results in the literature by showing that the Federbush theory is in fact completely ghost-free, even for a dynamical $U(1)$ field. Thus while the Federbush action describes a consistent effective field theory for the spin-2, it may be possible to extend the regime of validity of the theory above the scale $\Lambda_{q,3}$ provided one can make sense of the strong-coupling region. 
\\

A charged spin-2 field has several applications. It is known that Nature furnishes several composite charged massive spin-$2$ fields, e.g. hadronic resonances such as $\pi_2(1670)$, $\rho_3(1690)$, and $\al_4(2040)$.  Indeed, early attempts at bi-gravity and charged massive spin-2 fields were aimed at building a consistent description of these mesons \cite{Isham:1971gm}, and the work in constructing linearized charged spin-2 fields also existed to help describe mesons \cite{Federbush:1961}.

Additionally, a charged spin-2 field may be useful in condensed matter applications of the AdS/CFT correspondence, such as holographic superconductivity. For a review of holographic superconductors, see for example \cite{Hartnoll:2008kx}.
In studying superconductors, a standard set-up is to consider a black hole with scalar hair. The (massive) scalar field plays the role of spontaneously breaking a $U(1)$ symmetry, giving rise to superconductivity \cite{Weinberg:1986cq}. However, the scalar is only capable of describing S-wave superconductivity. In order to describe a D-wave superconductor, one needs black hole hair with charged helicity-2 dofs. A massive graviton can also be useful to break translation invariance in the bulk space, which can be useful for studying the DC conductivity. See also \cite{Vegh:2013sk, Blake:2013owa, Blake:2013bqa,Davison:2013jba,Zeng:2014uoa} for work on applying massive gravity in a holographic context.
\\

Especially in light of AdS/CFT applications, another question that we can ask is whether a charged spin-2 field can be consistently coupled to gravity. Given recent progress in massive gravity, one might hope that the key to describing gravitational interactions of a single, self-interacting, massive, charged spin-2 field will lie in the recently discovered non-linear ghost--free mass structure \cite{deRham:2010kj}.  It therefore seems timely to inspect if the recently discovered self-interacting massive spin-2 fields can help us describe a unitary Lagrangian for any sufficiently long-lived meson.\footnote{We do comment, however, that these mesons are resonances so their Lagrangians need only be effective field theories; thus unitarity may not be necessary as the finite lifetime of the resonance shows up as an imaginary part in the effective action.  Even still, in some limit one would na\"{i}vely expect that there ought to be a unitary theory of a single charged spin-2 field.}\\

Massive gravity  can be written as
\be
S = \frac{M_{\rm pl}^2}{2} \int \d^4 x \left[ \sqrt{-g} R - \frac{m^2}{2}\mathcal{U}\right]\,,
\ee
where the interaction potential $U$ is built out of a dynamical metric (with associated vielbein $e^a$) and a fixed reference metric (with associated vielbein $f^a$). The graviton potential that is free of the Boulware--Deser ghost at the non-linear is given by the set of interactions
\ba
\label{eq:drgt-interactions}
\mathcal{U}_1 &=& \ep_{abcd} e^a \wed e^b \wed e^c \wed f^d \nn \\
\mathcal{U}_2 &=& \ep_{abcd} e^a \wed e^b \wed f^c \wed f^d \nn \\
\mathcal{U}_3 &=& \ep_{abcd} e^a \wed f^b \wed f^c \wed f^d.
\ea
This form of the mass term was recently shown in \cite{deRham:2013awa} to emerge from an extra dimensional picture using Dimensional Deconstruction. Briefly, the Einstein-Hilbert term in 5 dimensions can be written in a particular gauge as
\be
S_{\rm G.R.,5d} = \int \ep_{abcd} \left(R^{ab} \wed e^c \wed e^d + \pa_y e^a \wed \pa_y e^b \wed e^c \wed e^d \right) \wed \d y,
\ee
where $y$ is a coordinate along the compact direction, and where we have temporarily neglected the zero modes corresponding to the radion and gravi--photon. We then discretize the compact direction, replacing the continuous coordinate $y$ by a discrete ``site index." In particular, by discretizing in the sense $\partial_y e^a_\mu \rightarrow m(e^a_{2}-e^a_1)$, we recover a particular combination of the ghost-free interactions in Equation~\eqref{eq:drgt-interactions}. By considering a more general discretization procedure, we may generate all of the interactions. This procedure was  also generalized to multi--gravity in \cite{Noller:2013yja}.

Deconstruction was also shown to be equivalent to truncating the Kaluza-Klein tower, essentially by interpreting $\pa_y \sim i n m$ for integer $n$. This suggests a method for generating a theory of a charged spin-2 field. In the Kaluza-Klein representation, the vielbeins are complex, $\tilde{e}^a_{n,\mu}$. In this representation the continuum theory has a global $U(1)$ symmetry under which
\be
\tilde{e}^a_{n,\mu} \rightarrow \tilde{e}^a_{n,\mu} e^{i n \theta}.
\ee
We may make this symmetry local by the minimal coupling replacement
\be
\d e \rightarrow D e = (\d - i q A \wedge)e.
\ee
In fact, the field $A$ appears naturally in the Kaluza-Klein context as a zero mode. In this context is sometimes known as the `gravi--photon.'

The $U(1)$ symmetry is associated with the group of continuous translations in the compact direction. This will be broken by a discrete subgroup upon discretizing. In Fourier space, this manifests itself by the presence of operators that violate charge conservation. However, there is a natural way to recover the $U(1)$ symmetry by simply projecting out those charge-violating operators. This will generate a candidate non-linear theory for a charged spin-2 field.

However, this projection modifies the kinetic structure. In light of recent results \cite{deRham:2013tfa}, we might expect that this will inevitably introduce ghosts. In fact we will give several arguments that the Boulware-Deser mode is present in any theory with a linearly realized $U(1)$.  Indeed, the new kinetic interactions we will derive by this method are closely related to the interactions considered in \cite{deRham:2013tfa}.\\

\noindent{\it Summary:} Our main results are
\begin{itemize}
\item The Federbush theory of a single massive spin-2 field interacting with a $U(1)$ gauge field, propagating on Minkowski space, is ghost-free. While it has been known that Federbush theory was ghost-free around constant electric field backgrounds (for example see \cite{Federbush:1961,Porrati:2010hm}), here we will present a proof that it is, in fact, fully ghost-free even with a dynamical $U(1)$ field. As a result, it is possible for the Federbush theory to have strongly coupled dynamics at a scale $q^{-1/3} m$ without violating unitarity.
\item There is a unique set of gravitational interactions for the Federbush theory in three dimensions that can be written with differential forms, and which reduces to Federbush around Minkowski space.
\item This unique non-linear extension to Federbush propagates a ghostly mode on a curved background.
\item As a by-product of our analysis, we develop some novel techniques to perform a constraint analysis based on \cite{Faddeev:1988qp} in the Einstein-Cartan formalism to check for the absence of ghosts. In Appendix \ref{app:bi-gravity}, we provide an alternative proof for the ghost freedom of bigravity in three dimensions (also known as Zwei-Dreibein gravity \cite{Bergshoeff:2013xma}) using our techniques.
\item We finally give an algebraic argument preventing the Einstein-Hilbert term from being compatible with a $U(1)$ symmetry.
\end{itemize}

\noindent {\it Outline:}
The rest of this work is organized as follows. In section~\ref{sec:Federbush} we shall review what is known about the linear theory of a charged, massive spin-2 field, and show that the Federbush action with a fully dynamical $U(1)$ gauge field is actually ghost-free.  In section~\ref{sec:charged-deconstruction},  we apply the method of Dimensional Deconstruction to generate a theory of a charged spin-2 field. In section~\ref{sec:non-minimal-extensions}, we shall show that generically actions that attempt to generalize Federbush to include gravitational coupling will introduce ghosts, by performing a constraint analysis in the \stu formalism. In section~\ref{sec:group-theory}, we analyze the group structure necessary for a non-linearly completed action, and demonstrate several of the initial problems with the theory. Finally, the appendices contain supplementary detail and alternative arguments.

\section{Ghost-free charged spin-2 fields on Minkowski}
\label{sec:Federbush}

Before attempting to construct an interacting theory of a massive charged spin-2 field, we first consider the flat space limit as a starting point. Federbush wrote down a theory of a single, massive charged spin-2 field in \cite{Federbush:1961} that was argued to propagate five dofs in the background of a constant electromagnetic field. Charged spin-2 fields have also been studied by Porrati in \cite{Porrati:2008gv}. In this section we will review what is known about the flat space case, following the discussion in \cite{Porrati:2008gv}. We will also find that the \textbf{Federbush theory} (also derived by Porrati) is completely \textbf{ghost free.} To our knowledge this goes beyond what has been done in the literature, where the stability analysis has been restricted to constant electromagnetic backgrounds.\\

We start with the Fierz-Pauli action for a complex spin-2 field $H_{\mu\nu}$
\be
\label{eq:flat-space-action}
S = \int \d ^4 x\  \left(H^*_{\mu\nu}\mathcal{E}^{\mu\nu\rho\sigma} H_{\rho\sigma} - m^2 \left( [H^* H] - [H^*] [H] \right) \right),
\ee
where $\mathcal{E}$ is the Lichnerowicz operator, normalized so that
\be
\E^{\mu\nu\rho\sigma} H_{\rho \sigma} = \ep^{\mu\rho \alpha \beta}\ep^{\nu \sigma \alpha'}_{\ \ \ \ \beta}\  \partial_\alpha \partial_{\alpha'} H_{\rho \sigma}= \square H^{\mu\nu} + \cdots
\ee
Square brackets refer to taking the trace with respect to the flat space-time metric, $[H] = \eta^{\mu\nu}H_{\mu\nu}$. Since $H_{\mu\nu}$ is complex, this theory propagates $2\times 5 = 10$ real dofs.

This theory has a global $U(1)$ symmetry under which $H\rightarrow He^{i\theta}$. We can make this symmetry local by coupling $H_{\mu\nu}$ to a $U(1)$ gauge field $A_\mu$, adding a kinetic term for $A_\mu$, and making the replacement
\be
\partial_\mu \rightarrow D_\mu = \partial_\mu - i q A_\mu,
\ee
where $q$ is the charge.

When applied to Fierz-Pauli, this procedure is ambiguous, because the covariant derivatives do not commute. When acting on a field $\phi$ with charge $q$,
\be
[D_\mu,D_\nu]\phi= - i q F_{\mu\nu}.
\ee
Since there are different representations of the Lichnerowicz operator that differ by integrating by parts and commuting partial derivatives, there are different ``minimal" covariantizations. The most general minimally coupled action is
\ba
\label{eq:minimally-coupled}
S = \int \d^4 x\ \Big(  \ep^{\mu\nu\rho\sigma}\ep^{\mu'\nu'\rho'}_{\ \ \ \ \ \sigma} &&  H^*_{\mu\mu'} D_\nu D_{\nu'} H_{\rho \rho'} - m^2 \left( [H^* H] - [H^*][H] \right) -\frac{1}{4} F_{\mu\nu}^2 \nn \\
&& +\  i q (2g-1) H^*_{\mu\nu}F^{\nu\rho} H_{\rho}^{\ \ \mu}\Big).
\ea
The ordering ambiguity is represented by the parameter $g$, which we may identify with the gyromagnetic ratio \cite{Deser:2001dt}. Already we may comment that from the point of view of an effective field theory, as long as this additional operator is not forbidden by some symmetry we expect it to arise, at least from quantum corrections.

Following \cite{Porrati:2008gv}, we can study this theory using a \stu analysis. We may introduce complex \stu fields
\be
\label{eq:charged-stuck}
H_{\mu\nu} = h_{\mu\nu} + D_{(\mu} \left( \frac{1}{m} B_{\nu)} + \frac{1}{2 m^2}D_{\nu)}\pi \right).
\ee
where $(a,b)\equiv ab+ba$, the action is invariant under charged linearized diffeomorphisms (diffs).

We now study the interactions in this theory, which arise entirely through the coupling between the $U(1)$ gauge field and the spin-2 field. That is, there are no self interactions of the spin-2 dofs.
It will be useful to consider a decoupling limit
\be
q\rightarrow 0, \ \ m\rightarrow 0, \ \ \Lambda_{q,n} \equiv \frac{m}{q^{1/n}}  \ {\rm fixed}.
\ee
The parameter $n$ will be fixed by the interaction that arises at the lowest scale in this limit. Interestingly for $q=m/\mpl$, $\Lambda_{q,n} = (m^{n-1} \Mpl)^{1/n}$, which we may identify as the usual scale $\Lambda_n$ arising in the effective field theory approach to massive gravity \cite{ArkaniHamed:2002sp}.

We may de-mix the kinetic term for the helicity-0 mode by performing the field redefinition
\be
h_{\mu\nu} \rightarrow h_{\mu\nu} + \frac{1}{2} \pi \eta_{\mu\nu}.
\ee
The kinetic terms for $h,B,\pi,A$ take the form
\be
S_{\rm kin} = \int \d^4 x \ \left(h^*_{\mu\nu}\E^{\mu\nu\rho\sigma}h_{\rho \sigma} - \frac{1}{4}|G_{\mu\nu}|^2 - \frac{3}{4} |\partial \pi|^2 - \frac{1}{4}F_{\mu\nu}^2\right),
\ee
where $G_{\mu\nu} \equiv \partial_\mu B_\nu - \partial_\nu B_\mu$

From  the scalings given in Equation~\eqref{eq:charged-stuck}, the kinetic terms for the \stu fields $B_\mu$ and $\pi$ do not scale with $q$.
Thus for a generic choice of $g$ we can identify the scale of the lowest order interactions as $\Lambda_{q,4}=q^{-1/4}m$. Explicitly the interactions are given by
\be
\mathcal{L}_{\Lambda_{q,4}} = (2g-1) \frac{i}{\Lambda_{q,4}^4} \partial_\mu \partial_\nu \pi^* F^{\nu\rho} \partial_\rho \partial^\mu \pi.
\ee
These interactions are higher derivative and signal the presence of ghosts arising at the scale $\Lambda_{q,4}$. Since this interacting is genuinely ghostly, we cannot imagine any strong-coupling self-unitarization mechanism to resolve it. Thus we may definitively say that the cutoff of this theory is at highest $\Lambda_c \sim \Lambda_{q,4}/(2g-1)^{1/4}$.

\subsection{Federbush is ghost-free}
However, as shown in \cite{Porrati:2008gv}, we may remove all interactions arising at the scale $\Lambda_{q,4}$ by the special choice of gyromagnetic ratio $g=1/2$ (this corresponds to the theory originally proposed by Federbush \cite{Federbush:1961}). In our conventions, it is clear that this choice corresponds to minimal coupling prescription
\be
\E^{\mu\nu}_{\mu'\nu'}\equiv \epsilon^{\mu\nu\rho\sigma}\epsilon_{\mu'\nu'\rho'\sigma} \partial_\rho\partial^{\rho'} \rightarrow  \epsilon^{\mu\nu\rho\sigma}\epsilon_{\mu'\nu'\rho'\sigma} D_\rho D^{\rho'}.
\ee
To identify the leading order interactions, we do an expansion in powers of $q$, keeping in mind that $D \sim \partial - q A$ and that $[D,D]\sim q F$. The leading order interactions come from the Lichnerowicz operator, which after introducing the \stu fields takes the schematic form
\be
\ep \ep \left(h^* + \frac{DB^*}{m} + \frac{D D \pi^*}{m^2} \right) D D \left( h + \frac{DB}{m} + \frac{D D \pi}{m^2} \right).
\ee
Let us first consider the interactions at order $q$. Because of the double epsilon structure the only non-vanishing term at this order uses the commutator to make $\ep \ep DDDB\sim q \ep \ep F \partial B$. The interaction arises at the scale $\Lambda_{q,3}$. It is given explicitly by
\be
\mathcal{L}_{\Lambda_{q,3}} = - \frac{i}{\Lambda_{q,3}^3} \epsilon^{\mu\nu\rho\sigma}\epsilon^{\mu'\nu'\rho'}_{\ \ \ \ \ \sigma}\partial_\mu \partial_{\mu'} \pi^* F_{\nu\rho} G_{\nu'\rho'} + c.c.\,.
\ee

Because of the double epsilon structure, the equations of motion for $\mathcal{L}_{\Lambda_{q,3}}$ are manifestly second order. As a result, the \textbf{Federbush theory is ghost-free at the scale $\Lambda_{q,3}$.} This means that there is no obstacle to treating the Federbush theory as a strongly coupled theory till energy scales $\Lambda_c$ where we could potentially have $\Lambda_c\gg \Lambda_{q,3}$, so long as no new dofs enter below $\Lambda_{c}$.

In fact, the Federbush theory is ghost-free to all orders in $q$. This follows directly from the double epsilon structure, which automatically removes any higher derivatives in the equations of motion. The ghost-freedom has also been explicitly checked by computing the equations of motion for the St{\"u}ckleberg--ed action and showing that all of the equations of motion are second order in time derivatives, using the techniques described in \cite{deRham:2013tfa}.

\subsection{Velo-Zwanziger problem in the \stu language}

Even though it is ghost-free, the Galileon-type structure of the interactions might lead us to suspect that the Federbush theory admits superluminal propagation around certain backgrounds. Indeed this is simply a manifestation of the well-known Velo--Zwanziger problem, expressed in modern language.

Let us consider an external electromagnetic field, $\bar{F}_{\mu\nu}$. Then the quadratic action for the perturbations is
\ba
S^{(2)} = \int \d^4 x\ \left( -\frac{1}{4} |G_{\mu\nu}|^2 - \frac{3}{4} |\partial \pi|^2 - \frac{i}{\Lambda_{q,3}^3}\left( \epsilon^{\mu \nu \alpha \lambda}\epsilon^{\rho \sigma \beta}_{\ \ \ \ \lambda} \partial_\nu \bar{F}_{\rho \sigma} \right) B^*_\mu \partial_\alpha \partial_\beta \pi + c.c. \right).\qquad
\ea
In this language it is clear that we can find backgrounds with superluminal group velocity. For example, perturbing around an electromagnetic background $\bar{F}_{\mu\nu}$, the operator $\Lambda_{q,e}^{-3} \partial \bar{F} B^* \partial^2 \pi \Lambda_{q,3}^3$ will modify the kinetic structure and can lead to superluminalities. This problem can occur even for arbitrarily small values of the electromagnetic field, since a sound speed $c_s^2 = 1+\epsilon$ for small $\epsilon$ is still superluminal.

In the literature the Velo--Zwanziger problem has traditionally been studied for backgrounds with a constant electromagnetic field. For such backgrounds, there is no contribution to the kinetic term at the scale $\Lambda_{q,3}$, as is evident by the expression above. Instead for background with constant electromagnetic fields, the leading correction to the kinetic term is schematically of the form
\be
\L_{{\rm int}, \Lambda_{q,2}}\supset\frac{1}{\Lambda_{q,2}^2}\bar{F} G^* G + \frac{1}{\Lambda_{q,2}^4} \bar{F}^2 \partial \pi^* \partial \pi.
\ee
Thus in standard presentations of the Velo-Zwanziger problem considering constant electromagnetic backgrounds, the superluminalities come from the operator that arises at a higher scale $\Lambda_{q,2}$.

We may tempted, as Velo and Zwanziger were, to attribute this apparent superluminality to a failure of causality. However the group and phase velocities can both be superluminal at low energies without conflicting causality, since the speed of information is set by the front velocity (see the review \cite{deRham:2014zqa} for a discussion and references on this point). The front velocity lies in the strong-coupling region for which this tree-level analysis is not appropriate. More precisely the test of causality is whether the commutator $[\pi(x),\pi(y)]$ vanishes outside the light cone. This vanishing is tied to the analyticity of its Fourier transform which is sensitive to the high energy behavior of the correlation function.
Group and phase velocities that exceed the speed of light in vacuum have been observed in nature (see for example \cite{2004PhRvL..93t3902B}). These measurements also explicitly confirm that the front velocity is luminal, consistent with causality. In addition it is known that the propagation of photons in a curved space-time can exhibit superluminalities in its low energy effective theory which are absent in the UV completion \cite{Hollowood:2007kt,Shore:2007um,Hollowood:2007ku,Hollowood:2008kq}.

\subsection{Ghost-free extensions to Federbush}
\label{sec:ghost-free-extensions}

We construct other charged spin-2 theories that are ghost free at the scale $\Lambda_{q,3}$ covariantizing the interactions proposed by Hinterbichler in \cite{Hinterbichler:2013eza}. For example, in $4+1$ dimensions we could have the operator
\be
\mathcal{L}_{kin}^{5d} = \frac{1}{\Lambda_{q,3}^{3}} \ep^{ABCDE}\ep^{A'B'C'D'E'}  H^*_{AA'} \(D_B D_{B'} H_{CC'}\) H^*_{DD'} H_{EE'},
\ee
where the capital indices $A,B=\{0,1,2,3,4\}$. The overall scale chosen so that a consistent decoupling limit exists at $\Lambda_{q,3}$. To identify the leading interaction, we may use the same argument as above, since all we have done is replacing one $\eta$ with an $h$. $\mathcal{L}_{kin}^{5d}$  gives rise to an interaction at $\Lambda_{q,3}$
\be
\mathcal{L}_{kin,\Lambda_{q,3}}^{5d} = \frac{i}{\Lambda_{q,3}^{6}} \ep^{ABCDE}\ep^{A'B'C'D'E'} \partial_A \partial_{A'} \pi^* F_{B'C'} G_{BC} h^*_{DD'} h_{EE'}.
\ee
As before, the double epsilon structure prevents higher order derivatives from appearing in the equations of motion. This can clearly be extended to the full set of interactions in any dimension, of the form $\ep \ep H^* DDH (H^*H)^{n}\eta^{d-6-2n}$ proposed in \cite{Hinterbichler:2013eza}. Thus, these represent consistent self-interactions of a spin-2 field on Minkowski space. 

However, in four dimensions there are no such terms invariant under a $U(1)$ symmetry, so we will not consider this possibility further in this work.

\subsection{First-order form}

It is useful to recast the Federbush action in first order-form. By first-order form, we mean that the action is written so that all fields appear with at most one derivative. This may be viewed as an intermediate-step in passing to the Hamiltonian. It will be convenient for us to work with first-order form when attempting to construct gravitational interactions.

To go to first-order form, we introduce a new field $\theta^{ab}_\mu$ (essentially the linearized spin connection) which plays the role of the momentum conjugate to $H$. We treat $\theta$ on equal footing as $H$. The first-order form for Fierz-Pauli is given by the action
\be
\label{eq:fierz-pauli-first-order-form}
S = \int \ep_{abcd} \left[ \( \d \theta^{ab} \wed H^{\star c} \wed \mathbf{1}^d + c.c.\) +2 \theta^{ae} \wed \theta^{\star eb} \wed \mathbf{1}^c \wed \mathbf{1}^d  \right],
\ee
where the one form $\mathbf{1}^a$ has components $\mathbf{1}^a_\mu = \delta^a_\mu$.

Upon integrating out the auxiliary field $\theta$, we find
\be
\theta^{ab}_\mu = -\frac{1}{2}\pa^{[a} H^{b]}_\mu.
\ee
In deriving this, we have used
\be
H_{a,\mu} = H_{\mu,a}.
\ee
This is the linearized version of the symmetric vielbein condition, which as is well-known is needed to show the equivalence of the vielbein and metric formulations of massive gravity.

Putting this back into the action (which is allowed since $\theta$ is not a dynamical field), we recover the usual form of the Fierz-Pauli action.

We can obtain a first-order representation of the Federbush action by simply following the minimal coupling procedure
\be
\d \theta \rightarrow D \theta^{ab} = (\d - i q A \wedge)\ \theta^{ab}.
\ee
Explicitly, the first order form for the Federbush action is
\be
\label{eq:federbush-first-order-form}
S = \int \ep_{abcd} \left[ \( D \theta^{ab} \wed H^{\star,c} \wed \mathbf{1}^d + c.c.\) +2 \theta^{ae} \wed \theta^{\star eb} \wed \mathbf{1}^c \wed \mathbf{1}^d \right].
\ee
A short calculation shows that integrating out $\theta$ reproduces the Federbush action.

It makes sense that covariantizing the theory in the first order form preserves the dofs. In first order form the dofs and Lagrange multipliers are manifest. We do not change the constraint structure by adding a gauge potential in this form.

Note that the spin connection is modified at the linear level due to the presence of the gauge field $A_\mu$. This behavior will persist at the non-linear level. Thus at the non-linear level it will be convenient to work in a first-order form.

\subsection{Gravitational interactions?}

\label{sec:non-linear-completion}

At this stage, from the point of view of massive gravity, the natural step is to try to construct a non-linear completion by adding self interactions for the graviton of the form $H^n, \partial^2 H^m$. The reason is that in the case of massive gravity, one expects to couple the massive spin-2 directly to the stress energy tensor $T_{\mu\nu}$ of matter fields. By the standard arguments (for example \cite{Feynman:1996kb}), this will force the spin-2 field to have non-linear interactions that realize a diffeomorphism symmetry.

However, as emphasized in the introduction, we do not have in mind that the charged spin-2 field is carrying a gravitational force. In other words, we will not couple charged spin-2 field to matter directly.
As a result, we do not necessarily need to add non-linear self-interactions to the massive graviton, beyond those considered in Section~\ref{sec:ghost-free-extensions}.

Nevertheless, it is interesting and important to understand the interactions of the massive charged spin-2 field with the true carrier of the gravitational force. In other words, we can view the charged spin-2 field itself as a matter field, and attempt to couple it to an electrically neutral, massless graviton.

In order to see if such an effective theory exists, we can get inspiration from multi-gravity and extra dimensional theories. A charged spin-2 field is built out of two real spin-2 fields. When we include the coupling to gravity, we will get a theory of multiple interacting spin-2 fields.
In fact, in section~\ref{sec:group-theory} we will show that it is impossible to view a charged spin-2 field as a gravitational theory with an Einstein-Hilbert kinetic term, because the charged spin-2 cannot realize a gravitational-type symmetry.\\

It is worth spending a moment defining what we would want for our gravitationally extended theory. We will require:
\begin{enumerate}

\item The gravitational extension should have a $U(1)$ symmetry. We will only consider non-linear extensions where this $U(1)$ symmetry is linearly realized.

\item The theory should be fully ghost free at all orders.

\item We would like a theory with a single charged spin-2 dof, coupled to gravity. As a result, the action should be built only out of a single complex spin-2 field $H_{\mu\nu}$, and a neutral metric $g_{\mu\nu}$.

\item The non-linear theory should reduce to the Federbush theory in the appropriate limit. Implicit in this requirement is that no dofs should become infinitely strongly coupled in this limit.
\end{enumerate}

\noindent Given these expected properties, we will attempt to construct a non-linear theory using various techniques. Ultimately we will discover that the ghost is re-introduced at some scale.

When considering non-linear completions, we will work in 2+1 dimensions. \textbf{We emphasize that a necessary condition for the theory to exist in higher dimensions is that it must work in 2+1 dimensions.}

This can be seen from multiple perspectives. If a consistent theory exists in 3+1 dimensions, there must be a consistent theory in 2+1 dimensions, because it is always possible to do a Kaluza-Klein compactification to reduce the 3+1 theory to the 2+1 theory. Furthermore, in $d$ spatial dimensions it is always possible to consider physical situations with translation invariance in $d-2$ spatial directions, so that the system effectively becomes $2+1$ dimensional. As a more general statement, there is no physical reason to expect that by adding more complication in extra spatial dimensions that we can resolve a difficulty that is already present in $2+1$ dimensions.\footnote{It is true that anomalies are strongly sensitive to the number of dimensions, however our main concern is the existence of the bosonic tree level theory which is largely insensitive to dimensions.}

One possible objection to this reasoning is that in higher dimensions we can add operators that would be topological in lower dimensions (such as the Lovelock terms), so there is more freedom in higher dimensions. In fact, as we will see, the main obstruction to constructing ghost-free theories of a charged spin-2 field is that the Einstein-Hilbert term itself is incompatible with the $U(1)$ symmetry. As a result we are forced to modify the kinetic structure, and this forces us to re-introduce the Boulware-Deser ghost. The higher order Lovelock terms will share this property. A group theoretic version of this argument, which is independent of spatial dimension, is given in section \ref{sec:group-theory}.

The main reason for working in 2+1 dimensions is that the theory in 2+1 dimensions is much easier to work with technically.  More detail on the advantages and formalism of 2+1 gravity (as well as conventions) are given in Appendix~\ref{app:3D-Gravity}. For other work studying the constraint analysis of massive-gravity type theories in three dimensions, see for example \cite{Banados:2013fda,Afshar:2014ffa,Alexandrov:2014oda}.

\section{Charged Deconstruction}
\label{sec:charged-deconstruction}
As described in the previous section, we will now be working in 2+1 dimensions for the remainder of the paper. Thus starting from this section, we will use Greek indices $\mu,\nu,\cdots$ to represent space-time indices in 2+1 dimensions. Capital Roman letters $M,N,\cdots$ will be used for space-time indices in 3+1 dimensions. In this section, we will also use a hat to distinguish between four dimensional exterior derivatives $\hat{\d}$ and three dimensional exterior derivatives $\d$.

We will apply the formalism of deconstruction to General Relativity in 3+1 dimensions to generate a candidate theory for a charged spin-2 field in 2+1 dimensions. First we will review the relevant Kaluza-Klein decomposition to clarify the gauge choices which are important for discretization. Then by discretizing the action we will generate a candidate theory for a massive graviton charged under a $U(1)$ group in 2+1 dimensions.

In fact, the na\"ive discretization process will break the $U(1)$ symmetry, because the continuous translation symmetry is broken to a discrete subgroup. However, we will find a natural way to restore the $U(1)$ symmetry in the resulting candidate theory. In the next sections, we will consider the consistency of this candidate non-linear extension.

\subsection{Kaluza-Klein with a vector zero mode}

As discussed in \cite{deRham:2013awa}, it is crucial to apply the deconstruction procedure using the vielbein language.\footnote{We will also work in the Euclidean, so all signature factors are $+1$ and the heights of Lorentz indices are not important.} The vielbein $E^A_M$ is related to the metric $g_{MN}$ by
\be
g_{MN} = E^A_M E^B_N \delta_{AB}.
\ee
For our purposes it will be useful to work with the Einstein-Cartan formalism, in which the spin connection $\Om^{AB}_{M}$ is treated as an independent variable that is determined by its own equation of motion. This is analogous to the Palatini formalism in the metric language.

In terms of $E$ and $\Om$, the 4 dimensional action for pure gravity is
\be
\label{eq:GR-4d}
S_{4d}[E,\Omega] =  \frac{\mpl^2}{4} \int \ep_{ABCD} R[\Omega]^{AB} \wed E^C \wed E^D,
\ee
where the Riemann curvature two-form is given by
\be
R[\Omega]^{AB}= \hat{\d} \Om^{AB} + \Om^{AC} \wed \Om^{CB}.
\ee
$S_{4d}$ is invariant under diffeomorphisms, under which $E$ and $\Om$ both transform as one-forms. It also enjoys a local Lorentz symmetry under which the fields transform as
\ba
\label{eq:llt-transformations}
E^A &\rightarrow& \Lambda^{AB}(x) E^B \nn \\
\Om^{AB} &\rightarrow& \Lambda^{AC} \Om^{CD} \Lambda^{DB} - \Lambda^{AC} \hat{\d} \Lambda^{CB}.
\ea
By  varying the action with respect to the spin connection one obtains the torsion-free condition in 4 dimensions
\be
\label{eq:4d-torsion-free}
\frac{\delta S_{4d}}{\delta \Om^{AB}_M} = 0 \implies  \hat{\d} E^A + \Om^{AB} \wedge E^B = 0,
\ee
and by varying with respect to the vielbein one obtains the vacuum Einstein equations
\be
\frac{\delta S_{4d}}{\delta E^A_M} = 0 \implies  \hat{\d} \Om^{AB} + \Om^{AC} \wed \Om^{CB} = R^{AB}(\Om) = 0.
\ee
We  perform a 3+1 split along the $y$ direction by parameterizing the vielbein as
\be
E^{A}_M \d x^M = \left(
\begin{array}{cc}
e^a_\mu \d x^\mu & N^a \d y \\
A_\mu \d x^\mu & N \d y
\end{array}
\right),
\ee
and the spin connections as
\be
\Om^{AB}_M \d x^M = \left(
\begin{array}{cc}
\om^{ab}_\mu \d x^\mu & \beta^{ab} \d y \\
K^a_\mu \d x^\mu & \lambda^a \d y
\end{array}
\right),
\ee
where $K_\mu^a \equiv \Om^{a4}_\mu$, $\lambda^a\equiv \Om^{a4}_y$.
In terms of these variables the action may be written as
\ba
\label{eq:S4d}
S_{4d} &=& \frac{\mpl^2}{4} \int \ep_{abc} \Big[ \left( R^{ab} - K^a \wed K^b \right) \wed e^c  \nn \\
&& + \left( \mathcal{D} \lambda^a - \partial_y K^a - \beta^{af} K^f \right) \wed e^b \wed e^c  \nn \\
&& + \left( \mathcal{D} \beta^{ab} - \pa_y \om^{ab} -  \lambda^{[a} K^{b]} \right) \wed A \wed e^c \Big] \wed \d y,
\ea
where $\ep_{abc}\equiv \ep_{abc4}$ and where $\mathcal{D}=\d + \om$ is the three-dimensional covariant exterior derivative and where $[a,b]=ab-ba$.

Our strategy will be to integrate out the components of the spin connection associated with the fourth direction, namely $\beta^{ab},K^a_\mu,\lambda^a$. The resulting action will be in a form appropriate for a three-dimensional observer, with a three dimensional spin connection $\om^{ab}$ and with fields transforming in the three-dimensional \poin group.
\\

First however we will fix some of the gauge freedom. This is an important step, because as discussed in \cite{deRham:2013awa}, different gauges in the continuum theory can produce different theories upon discretization.

\begin{itemize}

\item We fix 3 of the 6 Lorentz symmetries by setting
\be
N^a = 0.
\ee
We can do this by Lorentz transforming $N^a\rightarrow \Lambda^{ab} N^b + \Lambda^{a5}$ and taking $\Lambda^{a5}=-\Lambda^{ab}N^b$.

\item We also partially fix four of diff gauge symmetries by setting
\be
\partial_y A_\mu = 0, \ \ \ \partial_y N = 0.
\ee
We cannot use the gauge freedom to set $A_\mu=0$ and $N=1$ completely, we may only remove the $y$ dependence. These fields represent the massless zero vector and scalar modes.

\item In fact we will neglect the scalar mode (the radion). We are using Kaluza-Klein to motivate an action for a charged spin-2 field, and for these purposes the radion is not relevant. Thus we will set $N=1$ here.

\end{itemize}

With these gauge conditions in mind we can write down the torsion-free conditions \eqref{eq:4d-torsion-free} where at least one of the local Lorentz or space-time indices lie along the extra dimension
\ba
&&K^a_\mu = \partial_y e_\mu^a + \beta^{ab} e_\mu^b + \lambda^a A_\mu  \nn \\
&&F_{\mu\nu} = 2 K^a_{[\mu}e^a_{\nu]} \nn \\
&& \lambda^a e^a_\mu = 0.
\ea

These equations may be easily solved. The last equation sets $\lambda^a=0$ since $e^a_\mu$ is invertible. We may also take advantage of our remaining 3 local Lorentz gauge freedoms to set
\be
\label{eq:llt-gauge-choice}
\beta^{ab}= -\frac{1}{2} F^{ab} \equiv -\frac{1}{2} F^{\mu\nu}e_\mu^a e_\nu^b.
\ee
As a result of this gauge condition we may solve for $K^a_\mu$
\be
K^a_\mu = \partial_y e^a_\mu - \frac{1}{2}F^{ab} e^b_\mu.
\ee
The other equation of motion becomes
\be
\label{eq:symmetric-vielbein}
e^a_{[\mu} \partial_y e^a_{\nu]} = 0.
\ee
This last condition is the symmetric vielbein condition, here we see that it follows as an algebraic identity as a result of our gauge choice
\eqref{eq:llt-gauge-choice}.
\\

Finally plugging the solutions for the auxiliary fields $\beta,\lambda,K$ into the action\footnote{This is allowed since the equations of motion for these fields are algebraic.} \eqref{eq:S4d} we find
\ba
\label{eq:S4d-KK}
S_{4d} &=& \frac{\mpl^2}{2} \int \ep_{abc} \left( R^{ab} \wed e^c +  \pa_y e^a \wed \pa_y e^b \wed e^c  + 2 \pa_y \om^{ab} \wed A \wed e^c  \right)  \wed \d y \nn \\
&& + \frac{\mpl^2}{2} \int \d^3x \d y\  | e |\  \(-\frac{1}{4} F_{\mu\nu}F^{\mu\nu}\).
\ea
In deriving this expression we have used the fact that
\be
F^{af}e^f \wed \pa_y e^b \wed e^c \propto F_{\mu\nu} e^{\mu}_a \pa_y e^{\nu}_a =0,
\ee
using the symmetric vielbein condition~\eqref{eq:symmetric-vielbein}.

We also have set to zero an interaction
\be
\ep_{abc} F^{ab} A \wedge \mathcal{D} e^c.
\ee
We emphasize that we are not assuming the torsion-free condition $D e = 0$. We have not yet integrated out $\omega$. Based on the discussion in Section 2, we expect that the spin connection will be modified by the presence of the electromagnetic field. This implies that we do not want to assume the torsion free condition. However, $\mathcal{D} e \sim A$. As a result, after integrating out the spin connection, the above interaction will be proportional to $A \wedge A = 0$.

The presence of the Kaluza-Klein vector mode gives us a physically motivated starting point for considering theories of massive charged spin-2 fields. We will apply Dimensional Deconstruction to the four-dimensional action and generate a candidate action for a massive charged spin-2 field.

\subsection{Using deconstruction to generate a charged spin-2 theory}
Now we imagine a discrete set of $N$ special places along the fourth direction, with $y$ coordinate $y_I$, $I=1,\cdots,N$. We will now discretize the fourth compact dimension, keeping only the fields with located at $y=y_I$.
Following \cite{deRham:2013awa} we discretize the derivative $\pa_y$  in the sense
\be
\pa_y \phi(x^\mu, y_I) \rightarrow m \alpha_{IJ}\phi_J (x^\mu)\,,
\ee
where the $\alpha_{IJ}$ are in principle arbitrary coefficients that form some representation of a discretized derivative. Two natural choices considered in \cite{deRham:2013awa} are a ``local" discretization $\alpha_{IJ} = \delta_{I,J+1} - \delta_{I,J}$ and a ``truncated Kaluza-Klein" discretization $\alpha_{IJ} = \left[\sin(2\pi (I-J) / N)\right]^{-1}$.
We also replace the integral over $y$ with a sum over sites
\be
\int \d y f(y) \rightarrow \frac{1}{m} \sum_{I=1}^N f_I\,.
\ee
Applying this procedure to the action and canonically normalizing the photon kinetic term yields
\ba
\label{eq:charged-spin-2}
S_{3d} &=& \frac{M_3}{4} \int \ep_{abc} \sum_{I=1}^N \left( R(\omega_I)^{ab} + q \sum_J A \wedge \alpha_{IJ} \om_J^{ab} - m^2 \sum_{J,K} \alpha_{IJ}\alpha_{IK} e_J^a \wed e_K^b  \right) \wed e_I^c  \nn \\
&&  - \frac{1}{4} \int \d^3x \  | e |\  F_{\mu\nu}F^{\mu\nu}\,,
\ea
where $M_3 \equiv \mpl^2 / m$. We consider the charge $q$ to be an arbitrary parameter. The value of $q$ that arises from Deconstruction is
\be
q_{\rm Deconstruction} = \frac{m}{\sqrt{N} M_3} = \frac{m^2}{\sqrt{N} \mpl^2}.
\ee
The determinant $|e|$ and inverse vielbeins that appear in the photon kinetic term are somewhat ambiguous. In light of recent work on matter couplings \cite{Yamashita:2014fga,deRham:2014naa,deRham:2014fha}, the safest choice would be to have $|e|$ represent the determinant of a vielbein on just one site. In fact, in this work we will be mostly concerned with the self-interactions of the spin-2 field, and the determinant factor will not matter for the rest of our analysis.

\subsection{The Fourier transformed action}
\label{sec:fourier-transformed-action}
As discussed above, the discretized theory does not have a $U(1)$ symmetry. To see this it is easiest to set $q=0$ and to ignore the vector zero mode. We will then show that there is no global $U(1)$ symmetry present in this limit.

We may work in a representation where the (lack of) $U(1)$ symmetry is manifest by using a discrete Fourier transform
\be
\tilde{\Phi}_{n}^a = \frac{1}{\sqrt{N}} \sum_{I=1}^N \Phi_I^a e^{2 \pi i I n / N},
\ee
where $\Phi^a_I = \{ e^a_I, \om^{ab}_I \}$. Assuming $N$ is odd for simplicity, the inverse Fourier transform is then given by
\be
\label{eq:inverse-fourier}
\Phi_{I}^a = \frac{1}{\sqrt{N}} \sum_{n=-(N-1)/2}^{(N-1)/2} \tilde{\Phi}_n^a e^{-2\pi i I n / N }.
\ee
Note that while the $\Phi_I$ fields are real, the Fourier transformed fields $\tilde{\Phi}_n$ are complex. However since the $\tilde{\Phi}_n$ fields obey the condition $\tilde{\Phi}_n^* = \tilde{\Phi}_{-n}$, there are the same number of dofs in each representation (as there must be, since the discrete Fourier transform is an invertible field redefinition that cannot change the physics).

Interestingly, the $\tilde{\om}_n$ are not connections for $n \ne 0$. Instead, the $\tilde{\om}_n$ transform as tensors under diagonal local Lorentz transformations. To see this, note for example that in the case $N=3$ that
\be
\tilde{\om}_1 = \frac{1}{\sqrt{3}} \left( \frac{1}{2}(\om_3 - \om_1) + \frac{1}{2}(\om_3 - \om_2) + i \frac{\sqrt{3}}{2} (\om_1 - \om_2) \right).
\ee
Since the difference of two connections transforms as a tensor, $\tilde{\om}_1$ transforms as a tensor.

Treating the inverse discrete Fourier transform as a field redefinition, we may rewrite the action in the form
\ba
\label{eq:S3D-fourier}
S_{3d} &=& M_3 \int  \sum_I \Big[ \frac{1}{N} \sum_{n_1,n_2}  (\d \tilde{\om}^a_{n_1} \wed  \tilde{e}^a_{n_2}) e^{2\pi i I (n_1+n_2) / N} \nn \\
&+& \frac{1}{N^{3/2}}\sum_{n_1,n_2,n_3} \left(-\frac{1}{2} \ep_{abc}\tilde{\om}^a_{n_1} \wed  \tilde{\om}^b_{n_2} \wed  \tilde{e}^c_{n_3} \right) e^{2\pi i I (n_1 + n_2 + n_3) / N} \nn \\
&+& \frac{m^2}{N^{3/2}}\sum_{n_1,n_2,n_3} \left( \ep_{abc} \tilde{e}^a_{n_1}\wed \tilde{e}^b_{n_2}  \wed \tilde{e}^c_{n_3} \right) \left(  \sum_{J,K} \beta_{IJK} e^{\frac{2\pi i}{N} (I n_1 + J n_2 + K n_3) }\right)  \Big].
\ea
Here we have found it useful to define the coefficients $\beta_{IJK}$, instead of writing $\alpha_{IJ}\alpha_{IK}$.

\subsection{$U(1)$ symmetry in the $N\rightarrow \infty$ limit}
First let us consider the action in the continuum limit $N\rightarrow \infty$. In this limit the action returns to the Kaluza-Klein form \eqref{eq:S4d-KK} with $A=0$, and the discrete Fourier transforms become infinite Fourier transforms on $S^1$.

When Fourier transforming \eqref{eq:S4d-KK} we find
\ba
\lim_{N\rightarrow \infty}S_{3d} &=& M_4^2 \int_0^L \d y \int \Big[ \frac{1}{L} \sum_{n_1,n_2}  (\d \tilde{\om}^a_{n_1} \wed \tilde{e}^a_{n_2}) e^{2\pi i (n_1+n_2) x / L} \nn \\
&+& \frac{1}{L^{3/2}}\sum_{n_1,n_2,n_3} \left(-\frac{1}{2} \ep_{abc}\tilde{\om}^a_{n_1} \wed  \tilde{\om}^b_{n_2}  \wed \tilde{e}^c_{n_3} \right) e^{2\pi i (n_1 + n_2 + n_3) x / L} \nn \\
&+& \frac{1}{L^{3/2}}\sum_{n_1,n_2,n_3} (-n_1 n_2) \left( \ep_{abc} \tilde{e}^a_{n_1} \wed \tilde{e}^b_{n_2}  \wed \tilde{e}^c_{n_3} \right) e^{2\pi i (n_1 + n_2 + n_3) x / L }  \Big].
\ea
Then using the orthonormality relation
\be
\label{eq:orthonormality}
\int_0^L \d y e^{2 \pi i n y / L} = L \delta_{n,0},
\ee
we may perform the integrals over $y$ yielding
\ba
\lim_{N\rightarrow \infty} S_{3d} &=& M_4^2 \int \Big[ \sum_{n_1,n_2} \delta_{n_1+n_2,0}  (\d \tilde{\om}^a_{n_1} \wed  \tilde{e}^a_{n_2}) \nn \\
&+& \frac{1}{\sqrt{L}} \sum_{n_1,n_2,n_3} \delta_{n_1+n_2+n_3,0} \left(-\frac{1}{2} \ep_{abc}\tilde{\om}^a_{n_1}  \wed \tilde{\om}^b_{n_2} \wed  \tilde{e}^c_{n_3} \right) \nn \\
&+&\frac{1}{\sqrt{L}}\sum_{n_1,n_2,n_3} \delta_{n_1+n_2+n_3,0} (- n_1 n_2) \left( \ep_{abc} \tilde{e}^a_{n_1}  \wed \tilde{e}^b_{n_2} \wed  \tilde{e}^c_{n_3} \right) \Big].
\ea
In this form, it is clear that the theory has a global $U(1)$ symmetry because each interaction comes with a charge conserving delta function. This is nothing more than the usual statement that translation invariance in real space corresponds to momentum conservation in momentum space.

For our purposes, the physical significance of this observation is that if the global $U(1)$ symmetry were present, by introducing a gauge field $A_\mu$ and making the $U(1)$ symmetry local we could discover a theory of massive spin-2 particles charged under a $U(1)$ gauge symmetry. In the continuum four-dimensional theory, this minimally coupled field appears and is the massless KK vector mode. In the discretized theory, at finite $N$ it could in principle be any Abelian gauge field.

\subsection{Lack of $U(1)$ symmetry at finite $N$}
However the action \eqref{eq:S3D-fourier} does not have a global $U(1)$ symmetry at finite $N$. The reason is that the orthonormality relation \eqref{eq:orthonormality} is no longer valid. Instead, for integer $k$,
\be
\sum_{I=1}^N e^{2 \pi i I n / N} = N \delta_{n, k N} \ne \delta_{n,0}.
\ee
Applying this relationship, doing the sum over $I$ we arrive at the action
\ba
S&=& M_3 \int \Big[ \sum_{n_1,n_2} \delta_{n_1+n_2,0} (\d \tilde{\om}^a_{n_1}  \wed \tilde{e}^a_{n_2})\nn \\
&+& \frac{1}{\sqrt{N}} \sum_{k=-1}^1\sum_{n_1,n_2,n_3} \delta_{n_1+n_2+n_3,kN}\left(-\frac{1}{2} \ep_{abc}\tilde{\om}^a_{n_1} \wed  \tilde{\om}^b_{n_2} \wed \tilde{e}^c_{n_3} \right) \nn \\
&+& \frac{m^2}{N^{3/2}} \sum_{n_1,n_2,n_3}  \left( \ep_{abc} \tilde{e}^a_{n_1} \wed \tilde{e}^b_{n_2} \wed  \tilde{e}^c_{n_3} \right)  \left(  \sum_{I,J,K} \beta_{IJK} e^{\frac{2\pi i}{N} (I n_1 + J n_2 + K n_3) }\right) \Big].
\ea

When we truncate the sum at finite $N$, we must allow for operators that violate charge by an integer multiple of the number of fields. Thus the process of discretization breaks the $U(1)$ symmetry present in the continuum theory, corresponding to the statement that discretization has broken translation invariance in the compact direction.

Note that this subtlety does not affect the quadratic terms, because $n_1+n_2=k N$ implies $k=0$ for $|n| \leq (N-1)/2$. However charge violation is allowed for the cubic terms, because $n_1+n_2+n_3=k N$ implies $k=-1,0,1$. Thus the obstruction to the $U(1)$ symmetry arises only at the nonlinear level, and is invisible in the linear theory.

Explicitly, for $N=3$
\ba
S &=& M_3 \int  \ep_{abc} \Big( R[\tilde{\om}_0]^{ab} \wed \tilde{e}_0^c \nn \\
&& +\left[ (\d \tilde{\om}_1^{ab} + 2 \tilde{\om}_0^{ad} \wed \tilde{\om}_1^{cb})\wed \tilde{e}_1^{*,c} + c.c. \right] + \tilde{\om}^{ad}_1 \wed \tilde{\om}^{*,db}_1\wed \tilde{e}_0^c + m^2 \tilde{e}_1^a \wed \tilde{e}_1^{*,b}\wedge\tilde{e}_0^c \nn \\
&& + \left[\tilde{\om}_{1}^{ad} \wedge \tilde{\om}_1^{db} \wedge \tilde{e}_1^c + m^2 \tilde{e}_1^a \wedge \tilde{e}_1^b \wedge \tilde{e}_1^c + c.c. \right]  \Big).
\ea
This action violates global $U(1)$ invariance because of the interactions on the last line. The symmetry is broken both by the mass and kinetic terms.

\subsection{Restoring the $U(1)$ Symmetry for $N=3$}
Nevertheless, we may restore the $U(1)$. We simply introduce a projection operator that subtracts off the charge violating terms, by keeping only terms with $k=0$.

As we have seen, the quadratic terms are unaffected by this projection. The form structure implies that only cubic interactions can arise in three dimensions. Thus let us see what the impact of applying this projection operator is for a generic cubic interaction. We will specialize to the case $N=3$ for simplicity.

\subsubsection{Mass term}

The mass term is actually simpler to deal with. If we demand that the mass term has $U(1)$ invariance, we simply limit the permissible choices of $\beta_{IJK}$. Since the resulting mass term is still of the form of a ghost-free theory, there is no obstruction to choosing a $U(1)$ invariant mass term.

The $U(1)$ invariant mass term has two parameters
\be
\mathcal{L}_{m,U(1)} = m^2 \left( c_1 \tilde{e}_0 \wedge \tilde{e}_0 \wedge \tilde{e}_0 + c_2 \tilde{e}_1 \wedge \tilde{e}_1^* \wedge \tilde{e}_0\right).
\ee

Written in site language this amounts to a two parameter family for the $\beta_{IJK}$ coefficients
\ba
\label{eq:u1-mass-term-site-language}
\beta_{111} &=& c_1 + c_2 \nn \\
\beta_{112} &=& 3 c_1 \nn \\
\beta_{123} &=& 6c_1 - 3c_2,
\ea
with the rest of the $\beta$ coefficients determined by the various symmetries.
If we also impose the tadpole cancellation condition we are lead to the choice $c_1=1/2, c_2=3$, or
\ba
\beta_{111}=\frac{36}{5}, \ \ \beta_{112} = 3, \ \ \beta_{123} = -\frac{63}{5}.
\ea

\subsubsection{Kinetic term}

The kinetic term includes the cubic interaction
\be
S_{\rm cubic} = \frac{M_3}{3^{1/2}} \int \sum_{k=-1}^1\sum_{n_1,n_2,n_3} \delta_{n_1+n_2+n_3,kN}\left(-\frac{1}{2} \ep_{abc}\tilde{\om}^a_{n_1}  \wed \tilde{\om}^b_{n_2}  \wed \tilde{e}^c_{n_3} \right)\,.
\ee
If we focus on the terms with $k=\pm 1$ we find\footnote{For general odd $N$ we would need to write down one term for every triplet $(n_1,n_2,n_3)$ such that $|n_1|,|n_2|,|n_3| \leq (N-1)/2$ and $n_1 + n_2 + n_3 = \pm N$.  The number of solutions to these constraints increases with $N$, so $N=3$ is the simplest case.}
\ba
S_{\rm cubic}^{k=\pm 1} = -\frac{1}{2} \frac{M_3}{3^{1/2}} \int \ep_{abc} \left(\tilde{\om}_1^a \wed \tilde{\om}_1^b \wed \tilde{e}_1^c + \tilde{\om}_{-1}^a \wed \tilde{\om}_{-1}^b \wed \tilde{e}_{-1}^c \right)\,.
\ea
Performing the inverse Fourier transform \eqref{eq:inverse-fourier} yields
\be
S_{\rm cubic}^{k=\pm 1} = - M_3 \int \sum_{IJK} \gamma_{IJK} \ep_{abc} \om^a_I \wed \om^b_J\wed  e^c_K \,,
\ee
with
\be
\gamma_{IJK} \equiv \frac{1}{3^{2}}  \cos\left(\frac{2\pi }{3} (I + J + K)\right).
\ee
Thus in the site language, the $U(1)$ invariant cubic interaction takes the form
\ba
S_{\rm cubic}^{k=0} = S_3 - S_3^{k=\pm 1} = M_3 \int \sum_I \ep_{abc} \left[-\frac{1}{2} \om^a_I \wed \om^b_I \wed e^c_I +\sum_{J,K} \gamma_{IJK} \om^a_I \wed \om^b_J \wed e^c_K \right].\qquad
\ea
In other words, the cost of throwing out the terms that violate charge conservation in Fourier space is that we generate nonlocal terms upon taking the inverse Fourier transform. Thus we can write the full $U(1)$ invariant kinetic term as
\ba
S_{\rm kin}^{k=0} &=& S_{\rm GR} + S_{\rm kin}^{\rm new}\,,
\ea
where $S_{GR}$ is the sum of the usual Einstein Hilbert terms, and $S_{\rm kin}^{\rm new}$ is given by
\be
\label{eq:u1-kinetic-term-site-language}
S_{\rm kin}^{\rm new} = M_3 \int \sum_{IJK} \gamma_{IJK} \ep_{abc} \om^a_I \wed \om^b_J \wed e^c_K.
\ee
In this language, it is clear that \textbf{maintaining the $U(1)$ symmetry has forced us to modify the Einstein-Hilbert structure for the kinetic term.} As we will see in the next section, this is ultimately fatal.

In fact, the interactions we have generated are closely related to the interactions found by applying Dimensional Deconstruction to the Gauss-Bonnet term in 5 dimensions, as considered in \cite{deRham:2013tfa}. This can be made more explicit performing a field redefinition
\be
\om_1 \rightarrow \om_1 + (\om_2 - \om_3)
\ee
under which the Einstein-Hilbert term becomes
\be
R[\om_1]e_1 \rightarrow R[\om_1]e_1 + R[\om_2] e_1 + R[\om_3] e_1 - 2(\om_1 - \om_3)(\om_2 - \om_3)e_1.
\ee
The interactions $R[\om_1] \wedge e_2$ are of the same form as the interactions in \cite{deRham:2013tfa}. However these interactions are different because they are being considered in first-order form.

It is worth emphasizing that this illustrates again why the situation will not get better in higher space-time dimensions. By going to higher dimensions, we can potentially add more Lovelock terms. However the issue is that the Lovelock terms themselves necessarily break the $U(1)$ symmetry, and so must be modified. It is the modification to the Lovelock term that is ultimately responsible for re-introducing the Boulware Deser mode, as we will show in the next section. We have illustrated this explicitly in $2+1$ dimensions for the Einstein-Hilbert combination.

\subsection{Deconstruction-Motivated Charged Spin-2 Theory}
We have now reached the main result for this section, a natural candidate theory with a global $U(1)$ symmetry is
\ba
\label{eq:u1-theory-from-deconstruction}
S &=& S_{kin}^{k=0} + S_{mass}^{k=0} \nn \\
&=& M_3 \int  \ep_{abc} \Big( R[\tilde{\om}_0]^{ab} \wed \tilde{e}_0^c \nn \\
&& +\left[ (\d \tilde{\om}_1^{ab} + 2 \tilde{\om}_0^{ad} \wed \tilde{\om}_1^{cb})\wed \tilde{e}_1^{*,c} + c.c. \right] + \tilde{\om}^{ad}_1 \wed \tilde{\om}^{*,db}_1\wed \tilde{e}_0^c \nn \\
&& + m^2 \tilde{e}_1^a \wed \tilde{e}_1^{*,b}\wedge\tilde{e}_0^c \Big).
\ea
 A few remarks are in order:

\begin{itemize}

\item The next step, in principle, is to minimally couple a $U(1)$ gauge field through a minimal coupling procedure, $\d \rightarrow \d - i e A$. However, first we should check whether the candidate theory with a global symmetry is ghost free.

\item After introducing the gauge field through minimal coupling, the theory given in equation \eqref{eq:u1-theory-from-deconstruction} reduces to Federbush in the limit $M_3 \rightarrow \infty$. This is most easily seen by comparing the theory with the first-order form of Federbush given in \eqref{eq:federbush-first-order-form}.

\item Note that $S_{kin}^{new}$ has no dependence on the graviton mass $m$, the only scale present is $M_3$. This scale is completely fixed by the $U(1)$ invariance since $S_{kin}^{new}$ is not $U(1)$ invariant by itself, only the combination $S_{GR}+S_{kin}^{new}$ is $U(1)$ invariant.

\item $S_{kin}^{new}$ has diagonalized diff invariance, guaranteed by the form structure, as well as diagonalized local Lorentz invariance, which can be seen by expanding out the $\gamma_{IJK}$ explicitly
\be
S_{kin}^{new} = \frac{1}{9} M_3 \int \ep_{abc}\left[  2 (\om_1^a - \om_2^a)\wed (\om_1^b - \om_3^b) - (\om_2^a - \om_3^a)\wed(\om_2^b-\om_3^b) \right] \wed e_1^c + {\rm Z_3\ perms}.
\ee

\item We also see an advantage of working in the first order formalism. The equation of motion for the spin connections has been modified in a nontrivial way, and it is much easier to keep the spin connections as independent variables rather than needing to integrate them out explicitly.

\end{itemize}

\section{Degrees of freedom of generic non-linear completions}
\label{sec:non-minimal-extensions}

Rather than moving directly into establishing the number dofs of the action inspired by Deconstruction, we will now re-consider the problem of constructing a non-linear completion for Federbush from a more general perspective. The lesson from Deconstruction is that there is no way to associate a linearly realized $U(1)$ symmetry directly with the Einstein-Hilbert kinetic term. As a result, the first step is to try to find an appropriate ghost-free $U(1)$ invariant kinetic term for the spin-2 field.

As in the previous section, it is simpler to start by constructing a theory with no interaction with the $U(1)$ gauge field by taking the limit $q\rightarrow 0$. In this limit, the non-linear completion will have a global $U(1)$ symmetry. If the non-linear completion is ghost free for finite $q$, then the theory should also be ghost free in this limit.

We will write down the full set of terms in 2+1 dimensions consistent with the desired symmetries (a linearly-realized $U(1)$ symmetry). We will find a unique ansatz, which remarkably is equivalent to the one discovered using Deconstruction.

We will in fact show that there is no ghost-free, non-linear completion in three dimensions with a linearly realized global $U(1)$ symmetry. As a result, the corresponding theory with a local $U(1)$ with $q\ne 0$ cannot exist. Thus there is no non-linear ghost-free gravitational completion to Federbush.

\subsection{$U(1)$ invariant actions}

More precisely, let us start trying to build the most general non-linear theory, following the guidelines in section~\ref{sec:non-linear-completion}. The dofs should be limited to a single massive, charged spin-2 field $H^a_{\pm, \mu}$, and a dynamical vielbein $e^a_{\mu}$ that is neutral under the $U(1)$ symmetry representing a massless graviton.

We may always choose to work with a representation of the action where only first derivatives appear. We will choose to work with this form, to simplify the appearance of the non-linear interactions. In first order form, we also need to introduce auxiliary fields $\Theta_{\pm,\mu\nu}$ that carry information about the charged spin-2 fields.\\

The theory will be built out of the fields
\begin{itemize}
\item $e^a_\mu, \om^{ab}$, a gravitational background which transform as $U(1)$ scalars.
\item $H^a_{\pm,\mu}, \Theta^{a}_{\pm,\mu}$, which carry the charged spin-2 dofs. We take $H_- = H_+^*$. Under a $U(1)$ transformation with parameter $\alpha$, the spin-2 field $H_{\pm}$ transforms as $H_\pm \rightarrow e^{\pm i q \alpha} H_\pm$, and similarly for $\Theta_\pm$.
\end{itemize}

In terms of the language of the previous section, we may think of $\Theta^a$ as being the dual of the discrete Fourier transform of the spin connection, $\Theta_+^{a} = \ep^{abc} \tilde{\om}^{bc}_1$. However, here we are simply thinking of $\Theta^a_{\pm,\mu}$ as a field that will play the role of the momentum conjugate to $H^a_{\pm,\mu}$, without any a priori geometric interpretation (the fact that this can be done in a Lorentz-invariant way is what makes three dimensions special). Both $H$ and $\Theta$ transform as Lorentz and diff tensors.

$U(1)$ invariance is manifest in this representation. To ensure diagonal Lorentz invariance, the spin connection $\om$ should appear only through the curvature $R[\om]^{ab}$ or the exterior covariant derivative $\mathcal{D} = \d + \om$.\footnote{In principle since we are in three dimensions we could also add the gravitational Cherns-Simon term $\d \om \wedge \om + \om \wedge \om \wedge \om$, but we will not consider that possibility here.}

We will also limit our attention to actions that can be expressed in a wedge structure, without using a Hodge dual. We expect theories that are not of this form to have ghosts. As we will discuss in more detail below, in the \stu language, in order to avoid Boulware-Deser ghost modes it is crucial that some combination of the \stu fields are non-dynamical. However, for non-form like interactions this will almost always make the situation worse. The reason is that if we have a non-wedge interaction in unitary gauge
\be
H_{+,\mu}^a H_{-,\nu}^b X^{\mu\nu}_{ab}\,,
\ee
where $X^{\mu\nu}_{ab}$ is some function of the other fields. After introducing the \stu fields by $H = H + \mathcal{D} \phi$, this will have the form
\be
\mathcal{D}_\mu \phi^a \mathcal{D}_\nu \phi^b X^{\mu\nu}_{ab}\,,
\ee
which generically leads to kinetic terms for the \stu fields
\be
\dot{\phi}^a \dot{\phi}^b X^{00}_{ab}.
\ee
By local Lorentz invariance, this gives ALL of the \stu fields $\phi^a$ a kinetic term, and so they are all dynamical. This is already too many dofs, without even considering what happens to the Lorentz \stu fields, which either are also part of the momenta conjugate to $\phi^a$ or in principle could form their own dofs. The wedge structure will guarantee invariance under diagonal diffeomorphisms.

With these restrictions, the most general action, up to total boundary terms, is given by
\ba
\label{eq:wedgy-action-1}
S &=& M_3 \int  \ep_{abc} R[\om]^{ab} \wed e^c  \nn \\
&& + \left[ \( c_1 \mathcal{D} \Theta_{+}^{a} \wed H_{-}^a + c.c.\) + c_2 \mathcal{D} e^a_+ \wed e^a_{-} + c_3 \mathcal{D} \Theta_+^{a} \wed \Theta_{-}^{a}  \right] \nn \\
&& + \ep_{abc}\left( c_4 \Theta_{+}^{a} \wed \Theta_-^{b} \wed e^c + \(c_5 \Theta_{+}^{a} \wed H_-^{b} \wed e^c + c.c. \) \right) \nn \\
&& +\ep_{abc}\left( m^2 H_+^a \wed H_-^b \wed e^c + \Lambda\  e^a \wed e^b \wed e^c\right) .
\ea
Note that there is only one $U(1)$ invariant mass term when we separate out the cosmological constant, consistent with what was found above.

This action may be simplified with a field redefinition. We may factor the kinetic terms
\ba
\mathcal{L}_{kin} &=& \( c_1 \mathcal{D} \Theta_{+}^{a} \wed H_{-}^a + c.c. \) +c_2 \mathcal{D} e^a_+ \wed e^a_{-} + c_3 \mathcal{D} \Theta_+^{a} \wed \Theta_{-}^{a}  \nn \\
&=& c_2 \mathcal{D} \left( \Theta_+^a - C^{(+)} H^a_+ \right) \wedge \left( \Theta_-^a - C^{(-),*} H^a_-  \right) + c.c.\,,
\ea
where
\be
C^{(\pm)} = \frac{c_1}{c_2}\left(-1 \pm \sqrt{1-\frac{c_2 c_3}{|c_1|^2} } \right).
\ee
By performing a linear field redefinition,
\ba
\Theta_+^a - C^{(+)} H^a_+  & \rightarrow  & \Theta_+^a \nn \\
\Theta_+^a - C^{(-)} H^a_+  & \rightarrow & E_+^a,
\ea
while maintaining $\Theta_- = \Theta_+^*$ and $H_- = H_+^*$, we may set $c_2=c_3=0$.\footnote{The field redefinition is not invertible in the specific case when $|c_1|^2 = c_2 c_3$. However in that case the action is a perfect square and so after a field redefinition the action becomes $\mathcal{D} E_+ \wedge E_-$, so that $\Theta$ drops out of the kinetic term completely. We do not consider that case explicitly here since it does not reproduce Federbush and does not have the form of a charged spin-2 field.}
 This amounts to diagonalizing the kinetic term.

This linear field redefinition will of course renormalize the coefficients $c_4,c_5,m^2,\Lambda$, however since we have kept these parameters general up until now we will simply absorb the effects of the transformation into our definition of those parameters.
After this field redefinition, we may rescale the fields to absorb $c_1$ and $c_4$. Thus we are led to the action
\ba
\label{eq:wedgy-action-2}
S &=& M_3 \int \ep_{abc} R[\om]^{ab} \wed e^c  +  \( \mathcal{D} \Theta_{+}^{a} \wed H_{-}^a + cc\)  \nn \\
&& + \ep_{abc}\left(  \Theta_{+}^{a} \wed \Theta_-^{b} \wed e^c + \(c_5 \Theta_{+}^{a} \wed H_-^{b} \wed e^c + c.c. \) \right) \nn \\
&&+ \ep_{abc}\left( m^2 H_+^a \wed H_-^b \wed e^c + \Lambda\ e^a \wed e^b \wed e^c\right).
\ea
This is the most general non-linear completion, given the assumptions outlined above.

\subsubsection{Reproducing Federbush}
In fact, we may immediately conclude that $c_5=0$, just by the fact that having $c_5\ne 0$ does not reproduce the Federbush action at the linear level. Perturbing around flat space
\ba
&& e^a_\mu = \delta^a_\mu + \frac{1}{2\sqrt{M_3}}h^a_\mu \, ,\nn  \\
 && \om^{ab}_\mu = \frac{1}{\sqrt{M_3}}\theta^{ab}_\mu \, ,\nn \\
 && H_{\pm,\mu}^a = \frac{1}{2\sqrt{M_3}} h^a_{\pm, \mu} \, ,\nn \\
 && \Theta_{\pm,\mu}^a = \frac{1}{\sqrt{M_3}}\theta^a_{\pm,\mu},
\ea
and focusing only on the charged sector $h_{\pm},\theta_{\pm}$, the above action becomes
\be
S = \int \  \left( \d \theta_+^a \wed h_- ^a + c.c. \right) +  \ep_{abc} \theta_+^a \wedge \theta_-^b \wedge \mathbf{1}^c + \left( c_5 \ep_{abc} h_+^a \wedge \theta_-^b \wedge \mathbf{1}^c + c.c. \right).
\ee

Integrating out $\theta$, we find that
\be
\theta^a_{+,\mu} = \epsilon^{abc} \partial_b h^c_{+,\mu} + c_5 h^a_{+,\mu}.
\ee
Plugging this back into the action, we find the second order action
\be
S = S_{\rm F.P.} + \int \d^3 x \epsilon^{\mu\nu\rho} \partial_\mu h^a_{+,\nu} h^a_{-,\rho},
\ee
where $S_{\rm F.P.}$ is the usual Fierz-Pauli action.

Perhaps unsurprisingly, the non-Fierz-Pauli interaction has a ghost because the shift $h_{0i}$ appears with a time derivative, and so becomes dynamical. This may be seen in the \stu language as well. Replacing $h_{\pm,\mu\nu} \rightarrow h_{\pm,\mu\nu} + \partial_{(\mu} B_{\pm,\nu)}$ we find that
\be
\int \d^3 x \epsilon^{\mu\nu\rho}\partial_\mu \partial^\alpha B_{+,\nu} \partial_\alpha B_{-,\rho},
\ee
which has manifestly higher order equations of motion for $B_{\mu}$. This leads to a ghost in the free theory, so in the decoupling limit the ghost will be massless. This is unacceptable, so we conclude that $c_5=0$.

\subsubsection{Unique non-linear ansatz}
Thus we are lead to a unique ansatz for a the $U(1)$ invariant kinetic term
\ba
\label{eq:wedgy-action-3}
S &=& M_3 \int \ep_{abc} R[\om]^{ab} \wed e^c  + \( \mathcal{D} \Theta_{+}^{a} \wed H_{-}^a + cc\)  \nn \\
&& + \ep_{abc}\left(   \Theta_{+}^{a} \wed \Theta_-^{b} \wed e^c  + m^2 H_+^a \wed H_-^b \wed e_0^c + \Lambda e_0^a \wed e_0^b \wed e_0^c\right).
\ea
Remarkably, this is the action that we arrived at from the modified Deconstruction procedure in~\eqref{eq:u1-theory-from-deconstruction}, if we identify $H^a_\pm$ with $\tilde{e}^a_{\pm 1}$ and $\Theta^a_\pm$ with $\ep_{abc}\tilde{\om}^{bc}_{\pm 1}$.

We now want to determine the number of dofs in Equation~\eqref{eq:wedgy-action-3}. This can be done by an ADM analysis. However there is another way we can proceed, which we now describe.

\subsection{Phase space analysis of the non-linear theory}
We will do the analysis in the \stu language  directly in first-order form. The precise method we are using is new.

We will introduce \stu fields for the diffeomorphism and Lorentz symmetries. The advantage of this method is that all additional constraints other than the usual one which removes the BD ghost are first class. Then in principle one simply needs to count the dofs in the na\"ive phase space. This is sufficient to count the number of dofs, and thus we will be able to diagnose the presence or absence of Boulware-Deser modes.

In typical massive gravity and bi-gravity contexts, the analysis is done in second order form. In order to determine whether all the \stu fields are dynamical (in which case the Boulware-Deser ghost is present), one needs to check if the Hessian $\delta^2 S / \delta \dot{\phi}^a \delta \dot{\phi}^b$ is invertible (for example see \cite{deRham:2011rn}). However, this condition is extremely hard to check in the fully non-linear theory.

Nevertheless, there is an equivalent condition that we can use to simplify the analysis. If, and only if, the theory is free of the Boulware-Deser ghost, then the Boulware-Deser ghost mode should be absent in the quadratic lagrangian, perturbing around an arbitrary, off-shell background.

Thus we may diagnose the presence of a Boulware-Deser ghost by studying the quadratic action around an arbitrary, off-shell background. We can perturb the action in unitary gauge, and then introduce the \stu fields directly at the level of the perturbations. This greatly simplifies the way the \stu fields enter the action. Furthermore, it is much easier to establish the dofs of a quadratic action, than an arbitrary non-linear one.

The appendices contain some useful supplementary material. In Appendix~\ref{app:bi-gravity}, we apply this method to bi-gravity in three dimensions (also known as Zwei--Dreibein gravity) and confirm that bi-gravity is ghost free. In Appendix~\ref{app:flat-space-argument}, we perform a more brute force approach by perturbing to cubic order around Minkowski space.

\subsubsection{Strategy}
The starting point is to perturb the action around an arbitrary background
\ba
e^a_\mu &=& \bar{e}^a_\mu + h^a_\mu \nn \\
\om^{ab}_\mu &=& \bar{\om}^{ab} + \theta^{ab}_\mu \nn \\
H^a_{\pm,\mu} &=& \bar{H}^a_{\pm,\mu} + v^a_{\pm,\mu} \nn \\
\Theta^{a}_{\pm,\mu} &=& \bar{\Theta}^{a}_{\pm,\mu} + \mu^{a}_{\pm,\mu}.
\ea
As discussed above, we do not require the background to be on-shell.

We will then introduce the \stu fields directly at the level of the perturbations. Since we are dealing only with the quadratic action, we do not necessarily need to pattern the \stu fields off of the non-linear symmetry. It is enough to introduce enough new gauge symmetries to make all constraints first class, with corresponding phase space variables (i.e. we must introduce derivatives along with the fields). Additionally, we would like to maintain the background gauge symmetries (the diff and local Lorentz symmetries associated with the gravitational background $\bar{e}^a_\mu$) at the level of the perturbations. We will choose the following convenient \stu decomposition
\ba
v_{\pm}^a &\rightarrow& v^a_{\pm} + \bar{\mathcal{D}} \phi_{\pm}^a \nn \\
\mu_{\pm}^{a} &\rightarrow& \mu^{a}_{\pm} + \bar{\mathcal{D}} \lambda^{a}_{\pm}\,,
\ea
where $\bar{\mathcal{D}} \phi^a = \d \phi^a + \bar{\om}^{ab} \phi^b$ is the background covariant derivative.

Because we maintain the background symmetries, the action remains in first-order form after introducing the \stu fields. Terms with two derivatives can be rewritten as terms with one derivative on fluctuations after integration by parts. A generic term with scalar fluctuations $\chi^a$ and $\psi^b$ and a background field $\bar{\Phi}^a_\mu$ will have the form
\ba
\int \ep_{abc}\ \bar{\mathcal{D}} \chi^a \wed \bar{\mathcal{D}} \psi^b \wed \bar{\Phi}^c &=& \int \ep_{abc}\left(- \chi^a \bar{\mathcal{D}}^2 \psi^b \wed \bar{\Phi}^c + \chi^a \bar{\mathcal{D}} \psi^b \wed \bar{\mathcal{D}} \bar{\Phi}^c\right) \nn \\
&=& \int \ep_{abc}\left( -\chi^a \psi^d \bar{R}^{bd} \wed \bar{\Phi}^c + \chi^a \bar{\mathcal{D}} \psi^b \wed \bar{\mathcal{D}}\bar{\Phi}^c \right).
\ea
The antisymmetry of the wedge structure allows us to use the identity $\bar{\mathcal{D}}^2 \psi = \bar{R} \psi$.

Similarly, terms with three derivatives can be rewritten with one derivative using integration by parts and the Bianchi identity for the background, $\bar{\mathcal{D}} \bar{R} = 0$.
Additionally, it is clear that the zero components $h^a_0, \theta^{ab}_0, v^a_{\pm,0}, \mu^{a}_{\pm,0}$ will appear as Lagrange multipliers to this order because of the form structure.

The next step is to establish the size of the phase space. Before performing this step, we will first perform a counting argument to establish how Boulware-Deser ghost manifests itself in this representation.

\subsubsection{Degrees of freedom for healthy spin-$2$ fields in three-dimensions}
\label{sec:counting}

After introducing the diff \stu fields $B_I^\mu$ and Lorentz \stu fields $\lambda^{a}_I$, we may identify the dynamical fields and their conjugate momenta as follows:

\begin{itemize}
\item $\(e^a_{i}, \om^{ab}_i\)$: $6\ {\rm components} \times 2 = 12\ {\rm fields}$.
\item $\{H^a_{\pm,i},\Theta_{\pm,i}^a \}$: $6\ {\rm components}\times 2 \times 2= 24\ {\rm fields}$.
\item $\{\phi_{\pm}^a, \lambda^a_\pm \}$: ${3\ {\rm components}\times 2\ {\rm}}\times 2= 12\ {\rm fields}$.
\end{itemize}
We also have several first class constraints, associated with the gauge symmetries:
\begin{itemize}
\item $3$ diagonal diffeomorphism symmetries (with Lagrange multipliers $e_0^a$).
\item $2\times 3$ \stu diffeomorphism symmetries (with Lagrange multipliers $H_{\pm,0}^a$).
\item $3$ local Lorentz symmetries (with Lagrange multipliers $\om_{0}^{ab}$).
\item $2 \times 3$ \stu local Lorentz symmetries (with Lagrange multipliers $\Theta_{\pm,0}^a$).
\end{itemize}
Thus, in general the dof counting is
\ba
&& (12+24+12)\ {\rm dynamical\ variables}\nn \\
&& - 2 \times 18\ {\rm first\ class\ constraints}  \nn \\
&&= 2 \times (2+2) + 2\times (1+1)\ {\rm dofs}.
\ea
A massless graviton has $0$ propagating dofs in three dimensions, and a charged massive graviton has $2 \times 2=4$. So we expect to have $8$ phase space dofs. These $8$ phase space dofs are represented by the first term above. The second term represents $2$ extra phase space dofs for each of the St\"uckelbergized sites. This corresponds to one extra scalar dof for each of the massive modes, which is the usual Boulware-Deser ghost.

In order to avoid the existence of these Boulware-Deser ghost modes, we must project out some of the phase space dofs. This must be done by writing an action where four independent combinations of the \stu fields are non-dynamical.\footnote{In principle one could imagine adding second class constraints by hand to remove the Boulware-Deser ghost. However these constraints would need to be Lorentz invariant. We do not consider this possibility likely.}

\subsubsection{Constant gravitational background}

Having set up this formalism it is not hard to see that there is a ghost. We simply need to work on a fixed gravitational background, with $h=\theta=0$. In order to avoid a ghost, it is necessary for the theory to be ghost-free with a fixed gravitational background. We will also assume the background is torsion free, $\bar{\mathcal{D}} \bar{e} = 0$.

Perturbing our non-linear ansatz \eqref{eq:wedgy-action-3} around an arbitrary off-shell background, and introducing the \stu fields, we are led to
\be
S = \int \bar{\mathcal{D}} \mu^a_+ \wed \left[v^a_- - \ep_{abc} \lambda^b_- \bar{e}^c \right] + \bar{\mathcal{D}} \phi_+^a \wed \left[  \bar{R}^{ab} \lambda^b_- + m^2 \ep_{abc} v_-^b \bar{e}^c \right] + c.c. + {\rm N.D.}
\ee
where ${\rm N.D.}$ refers to terms with no derivatives acting on the fluctuations or \stu fields.

Focusing on time derivatives this becomes
\be
S = \int \d^3 x \ \dot{\mu}^{+,a}_i P^{-,a}_i + \dot{\phi}^{+,a} \pi^{-,a} + c.c.
\ee
where
\ba
P^{-,a}_i &=& \ep_{ij}v^{-,a}_i - \ep^{abc} \lambda^{-,b} e^{0,c}_j \nn \\
\pi^{-,a} &=& \ep^{abc} \ep^{ij} \lambda^{-,b} R^{c}_{ij} + m^2 \ep^{abc} \ep_{ij} v^{-,b}_i e^{0,c}_j.
\ea
Note that in first-order form the Lorentz \stu fields $\lambda^a$ play the role of momenta conjugate to the diff \stu fields $\phi^a$. This is explored in more detail in Appendix~\ref{app:flat-space-argument}.

The key issue is whether or not all of the \stu fields have independent conjugate momenta. We can make this more explicit by rewriting $\pi^{-,a}$ in terms of $P^{-,a}$

\be
\pi^{-,a} = \ep^{abc} \ep_{ij}  \left[  \left(\bar{R}^b_{ij} + m^2 \ep^{bpq} \bar{e}^{p}_i \bar{e}^{q}_j \right) \lambda^{-,c} + m^2 P^{-,b}_i \bar{e}^{c}_j \right].
\ee
The worrying term is the first term, proportional to $\lambda^{-,c}$. The reason is that $P^{-,a}_i$ is already a momentum conjugate to $e^{+,a}_i$, so if one linear combination of the $\pi^{-,a}$ depends only $\sim P^{-}$ then there is no \textbf{independent} momentum for the corresponding linear combination of the $\phi^{-,a}$. A different version of this argument is given in Appendix~\ref{app:symplectic-form}.

For Minkowski space, with $\bar{R}=0$ and $\bar{e}^a_i = \delta^a_i$, we find that the \stu field $\phi^{+,0}$ does not have an independent conjugate momentum, because $\pi^{-,0}=m^2 P^{-,i}_i$. This is simply a confirmation in three dimensions of the fact that the first order form of Fierz-Pauli is ghost-free.

However for a generic background, all three components of $\pi$ will be independent of $P$ through the dependence on $\lambda$. Thus around curved backgrounds, the Boulware-Deser mode will appear in the phase space.

To summarize, we have shown that the unique form-like extension of the Federbush theory contains a Boulware-Deser mode in the $q\rightarrow 0$ limit. The argument in this section covers both of the possible kinds of non-linear completion discussed in section~\ref{sec:non-linear-completion}.  Since the Federbush theory propagates ten dofs, the kinetic term of the ghost vanishes around Minkowski, so the new dof if taken seriously would be infinitely strongly coupled around Minkowski space. However, from an effective field theory point of view, the new dof can be taken as an indication of a higher derivative terms in the Lagrangian which indicates unitarity violation at some scale. As usual in an EFT as long as we consider physics below that scale then the ghostly mode can be harmless. The crucial point is that the scale of the ghost is hierarchically above the strong-coupling scale of the Federbush theory.

\section{Group-theoretic obstructions to non-linear charged spin-2 fields}
\label{sec:group-theory}

Having demonstrated the general problems associated to our attempt to enforce $U(1)$ symmetry on spin-2 fields, we now present an alternative argument. While before we focused on specific lagrangians, and found it easier to work in $2+1$ space-time dimensions, in this section we will give a group theoretic argument that works in any space-time dimension. Thus in this section we will work in $d+1$ space-time dimensions, and show that there is a group theoretic explanation for why it is impossible to construct a $U(1)$ invariant theory while preserving $ISO(1,d)$ symmetry associated to the spin-2 field in $d+1$ space-time dimensions.

\subsection{Obstructions to finding $[ISO(1,d) \times ISO(1,d)] \rtimes U(1)$}

If we suppose that the kinetic terms must be given by an Einstein-Hilbert kinetic term, then this entails two distinct copies of the $ISO(1,d)$ algebra, one for each copy of the Einstein-Cartan action.  Then the two vielbein together form massive representations of the \poin group after the ghost-free mass terms are added.  The mass terms breaks one copy of the local $ISO(1,d)$ symmetries.  This copy can be restored via a set of \stu fields, this makes the symmetry $ISO(1,d) \times ISO(1,d)$ non-linearly realized, but still present.  The $U(1) \cong SO(2)$, contrariwise, must mix with these two \poin algebras.  This is because at the level of field representations (using, for the moment, the real representation of the $2$ of $SO(2)$), we see that
\be
\de_{U(1)}(\theta)E^a = \theta \ep_{ij} E_j^a \Longrightarrow [Q,P^a_i] = \ep_{ij} P^a_j\,.
\ee
This is because when one says that a spin-$j$ particle is ``charged'', one means that the particle is both complex (in other words, the $2$ of $U(1)$) and a spin-$j$ representation of the \poin group.  This tells us, then, that the group, $G$, that we are looking for is of the form
\be
		G = U(1) \rtimes [ ISO(1,d) \times ISO(1,d) ]\,.
\ee
It is natural then to ask if one can consistently construct this group.  We shall assign $Q$ as the generator of $U(1)$ and ${P^a_i, M^{ab}_i}$ as the generators of $ISO(1,d) \times ISO(1,d)$. If we attempt to construct
the given algebra, we find that the following follows without issue
\ba
	&& [P^a_i,\, P^b_j] = 0 \nn \\
	\,&& [Q, \, P^a_i] = \ep_{ij} P^a_j\\
	\,&& [Q, \, M^{ab}_i] = \ep_{ij} M^{ab}_j\,.
\ea
Unfortunately, a problem arises for the following commutation relations
\ba
	[P^a_i,\, M^{bc}_j] &=& \text{?} \\
	\, [M^{ab}_i, \, M^{cd}_j] &=& \text{?}
\ea
The issue here is that the two algebras should separately generate two copies of $ISO(1,d)$.  However, the $SO(2)$ index is clearly obstructing this, since the exact object we would need in order to accomplish this would a structure constant, $f^{ijk}$, in order to work correctly (i.e. convert two indices into one free index).
Unfortunately, it is well known that $U(1)$ is abelian, and thus $f^{ijk} \equiv 0$, and thus we see that because the $U(1)$ is abelian, this requires that the generator $M^{ab}_i$ commute with all other generators.  In other words, the non-Abelian properties of generators are incompatible with the $2$ of $U(1)$ structure.

This can be seen even at the level of Yang-Mills, for reference, where the non-Abelian internal group G cannot be semi-direct producted into the internal group (i.e. there cannot be a self-charged photon under an abelian symmetry); a consistent theory can only be made with a direct product.\footnote{
In the Standard Model, there is a $U(1)_Y$ for the hypercharge structure, but this generator does not mix the non-Abelian $SU(2)_L \times SU(3)_C$, keeping it self-consistent.}

\subsection{Checking the Jacobi identity}

One may also see this by analyzing the Jacobi identity. Here we write the most natural commutation relations to force the $P$ into a $2$ of $U(1)$.  The established commutation relations for the generators $\{ Q, M^{ab}_1, M^{ab}_2, P^c_1, P^c_2 \}$ is the following:
\ba
    && [Q,Q] = 0 \nn \\
	\,&& [P^a_1,\, P^b_1] = [P^a_1,\, P^b_2] = 0 \,\,\,\,\,\,\,\,\,\,\,\,\,\,\,\,\,\,\,\,\,\,\,\,\,\,\,\,\,\,\,\,\,\,\,\,\,\,\,\,\,\,\,\,\,\,\,\,\,\,\,\,\,\,\,\,\,\,\,\,\,  \text{ (Same for 1 $\leftrightarrow$ 2) }\nn \\
	\,&& [M^{ab}_1,\,P^c_2] = [M^{ab}_1,\,M^{cd}_2] =  0 \,\,\,\,\,\,\,\,\,\,\,\,\,\,\,\,\,\,\,\,\,\,\,\,\,\,\,\,\,\,\,\,\,\,\,\,\,\,\,\,\,\,\,\,\,\,\,\,\,\, \text{  (Same for 1 $\leftrightarrow$ 2) } \nn \\
	\,&& [M^{ab}_1,\,P^c_1] = \eta^{bc} P^a_1 - \eta^{ac} P^b_1 \hspace{130pt} \text{ (Same for 2) } \nn \\
	\,&& [M^{ab}_1,\, M^{cd}_1] = \eta^{ac}M^{bd}_1 +  \eta^{bd}M^{ac}_1 -\eta^{ad}M^{bc}_1 - \eta^{bc}M^{ad}_1. \,\, \text{(Same for 2)} 
\ea
However, the following Jacobi identity can be seen to fail:
\ba
	&& \Big[Q, [M_1,M_2]\Big] + \Big[M_1,[M_2,Q]\Big]  + \Big[ M_2, [Q,M_1] \Big]  \nn \\
	&&= \Big[Q,0 \Big] + \Big[M_1,M_1 \Big] + \Big[ M_2, M_2 \Big] \nn \\
	&&= \eta^{ac}\( M^{bd}_1 + M^{bd}_2\) +  \eta^{bd}\(M^{ac}_1 + M^{ac}_1 \) -\eta^{ad}\(M^{bc}_1 - M^{bc}_2\) - \eta^{bc}\(M^{ad}_1 + M^{ad}_2\) \nn \\
	&& \ne  0,
\ea
and thus these generators fail to form a Lie algebra, which means that exponentiating them will fail to lead to a closed Lie group. 

\subsection{Kac-Moody algebra admits no finite truncations}

Finally, a different approach can be found by studying Kac-Moody algebras, see Ref.~ \cite{Dolan:1984fm}.  If one takes the usual prescription for Kaluza-Klein compactification (here done in a different gauge, but a separate gauge-fixing will result in the same story), where the gauge fields are parameterized as follows
\ba
	g_{MN} = \phi^{-1/3} \begin{pmatrix}
															g_{\mu\nu} + \mpl^{-2}\phi A_{\mu}A_{\nu} && \mpl^{-1}\phi A_{\mu} \\
															\mpl^{-1}\phi A_{\nu} && \phi
											 \end{pmatrix}.
\ea
Then after performing the Fourier expansion over the compact extra dimension, $y \in [0,2\pi[$, we find the infinite tower of modes
\ba
\label{eq:KK_expan}
	&& g_{\mu\nu}(x,y) = \sum^{\infty}_{n=-\infty} g_{n}\,_{\mu\nu}(x) e^{iny} \, ,  \nn \\
	&& A_{\mu}(x,y) = \sum^{\infty}_{n=-\infty} A_{n}\,_{\mu}(x) e^{iny} \, , \nn \\
	&& \phi(x,y) = \sum^{\infty}_{n=-\infty} \phi_{n}(x) e^{iny}.
\ea
To see where the Kac-Moody algebra derives from (and thus that the 4-D field content forms a representation of a Kac-Moody algebra), we may take the $D=5$ \poin algebra made of generators $\hat{P}_{M}$ and $\hat{M}_{MN}$
\ba
	&& [\hat{P}^M,\, \hat{P}^N] = 0 \, \nn \\
	\,&& [\hat{M}^{MN},\,\hat{P}^R] = \eta^{NR} \hat{P}^M - \eta^{MR} \hat{P}^N \,  \nn \\
	\,&& [\hat{M}^{MN},\, \hat{M}^{RS}] = \eta^{MR}\hat{M}^{NS} +  \eta^{NS}\hat{M}^{MR} -\eta^{MS}\hat{M}^{NR} - \eta^{NR}\hat{M}^{MS}.
\ea
Performing a $(4+1)$-split on these generators, where $\mu = 1, \, 2, \text{ and } 3$ and $x^5 \equiv y$, we have the following form\footnote{We have explicitly broken $\hat{M}^{5\mu}$ by our choice of global topology, $R^{1,3}\times S^1$.}
\ba
	&& \hat{P}^{\mu}(x,y) \to  P_n\,^{\mu} = e^{iny}\p^{\mu} \, ,  \\
	&& \hat{P}^5(x,y) \to   M_n\,^{\mu\nu} = ie^{iny} \p^y \,, \\
	&& \hat{M}^{\mu\nu}(x,y) \to  Q_n = e^{iny} \big( x^{\nu}\p^{\mu} - x^{\mu}\p^{\nu} \big).
\ea
Next we impose the conditions \eqref{eq:KK_expan} on this splitting, which yields the following Kac-Moody algebra
\ba
		&& [P_n\,^{\mu},\, P_m\,^{\nu}] = 0 \, , \nn \\
	\,&& [Q_n,\, P_m\,^{\mu}] = -m P_{m+n}\,^{\mu} \, , \nn \\
	\,&& [Q_n,\, Q_m] = (n-m) Q_{m+n} \, , \nn \\
	\,&& [M_n\,^{\mu\nu}, \, P_m\,^{\rho}] = \eta^{\nu\rho} P_{n+m}\,^{\mu} - \eta^{\mu\rho} P_{n+m}\,^{\nu} \,  ,\nn \\
	\,&& [M_n\,^{\mu\nu}, \, Q_m ] = -m M_{m+n}\,^{\mu\nu} \, , \nn \\
	\,&& [M^{\mu\nu},\, M^{\rho\si}] = \eta^{\mu\rho}M_{m+n}\,^{\nu\si} +  \eta^{\nu\si}M_{m+n}\,^{\mu\rho} -\eta^{\mu\si}M_{m+n}\,^{\nu\rho} - \eta^{\nu\rho}M_{m+n}\,^{\mu\si}\,,
\ea
of which there are several important things to note.  Firstly, this is an infinite-dimensional Lie algebra, since the index $n$ on the generators runs over all integers.  The second thing to note is that there
cannot be a finite truncation of the generators containing multiple copies of the \poin generators.The algebra only consistently closes, for instance, when the Lorentz generators form their Virosoro-like algebra.

This is group-theoretic explanation for why one can have linearly realized charged spin-2 fields if there is an infinite tower, but finite truncations are inconsistent.  In other words, this is why Dimensional Deconstruction recovers a copy of $U(1)$ as $N\to \infty$.  We see here that we cannot simultaneously diagonalize the charge basis and the Lorentz boost or momentum basis (since they do not commute), and thus charge always entangles itself into these operators. The only exception, of course, is if we take a finite truncation of a single graviton, but this prohibits us from having $1<N<\infty$ number of gravitons.  This means that we can see the failure to generate consistent charge spin-2 theories from dimensional deconstruction's relationship with the Kaluza-Klein procedure.

Ostensibly, one might expect that there could be an alternative infinite-dimensional algebra that one might generate by sending $N\to \infty$ in Dimensional Deconstruction (with a different topology, for instance), which might have a consistent finite truncation.  However, this is why the non-existence of the group $[ISO(1,d) \times ISO(1,d)] \rtimes U(1)$ will prevent this from happening.  The most one can hope for is $[ISO(1,d) \times ISO(1,d)] \times U(1)$. These group theoretic arguments appear to be consistent with our explicit findings. 

\section{Discussion}

We have explored whether the ghost-free properties of massive gravity might allow for the existence of a single charged spin-2 field.  We have defined a set of natural requirements for a charged spin-2 field:

\begin{itemize}
\item[1.)] A ghost-free theory of a single massive complex spin-2 field with a linearly realized $U(1)$ symmetry.
\item[2.)] This theory simultaneously exhibits a non-linearly realized double copy of $ISO(1,d)$ symmetry through its \stu fields. (Or triple copy if a massless spin-2 field is added).
\end{itemize}

Using a modified variant of Dimensional Deconstruction that unfreezes the vector zero-mode of the graviton (i.e. the gravi--photon), we obtained an interesting model that had many novel and non-trivial features such as spin-1 and spin-2 coupling, but ultimately broke the $U(1)$ symmetry.  We see that through a straightforward process, the $U(1)$ symmetry may be restored to a unique theory.  This unique candidate theory has manifest $U(1)$ invariance, and can be derived from only assuming $U(1)$ invariance and a general form structure.  Unfortunately, the $U(1)$ structure explicitly breaks the finely-tuned structure of the kinetic term for General Relativity, and the de-tuning was demonstrated to give rise to a spurious BD ghost dof which is infinitely strongly coupled around flat-space. This more or less prohibits such a theory arising in higher dimensions, since they would presumably have to give rise to a healthy three-dimensional theory via dimensional reduction. From an EFT point of view the existence of the BD ghost may just be taken as an indication of higher derivative operators in the EFT. These operators will be suppressed by a scale which tends to infinity in the limit $\mpl \rightarrow \infty$ in which we recover the Federbush theory. 

Alternatively, one can view this question from the standpoint of group theory, as was done previously \cite{Dolan:1984fm}.  Without assuming any higher dimensional structure, we explicitly demonstrate that there cannot exist a group mixing the vielbein (and thus the copies of $ISO(1,d)$), because doing so requires a violation of the Jacobi identity and the group cannot close.  In essence, this is the obstruction to the theory that was almost generated by Dimensional Deconstruction, where the $U(1)$ leaves the action invariant and the algebra closes only when the number of gravitons is taken to infinity, at least one such example of a resulting consistent is that of Kac-Moody.  It is previously well-known that Kac-Moody has no finite subgroup containing two or more copies of the \poin group. In principle, one might imagine that the infinite collection of gravitons might give rise to other infinite-dimensional Lie algebras that could, and therefore it is useful to see explicitly that the guilty assumption lies in $U(1)$ rotating the \poin copies into one another, and thus this gives a rather general argument against such a structure.

However, if one weakens this requirement, as is done in the unique candidate theory, and instead relies on not making the $ISO(1,2)$ symmetry manifest, one enforces the $U(1)$ symmetry from the outset, it breaks the semi-direct product into a direct product.  Again, the resulting theory appears to be unique, assuming that it can be cast into differential form, but it gives rise to an unphysical dof.

Nevertheless we attempted to construct an appropriate $U(1)$ invariant kinetic term that was not of the Einstein-Hilbert form. We showed that the new kinetic term that we created propagated a Boulware-Deser ghost by performing a \stu analysis directly in first-order form. The methods described in this paper can be extended easily in three dimensions to discuss the first-order form of the kinetic interactions described in \cite{deRham:2013tfa}. It would also be interesting to extend this method to four dimensions, however this is complicated by the well-known fact that in dimensions greater than 3 the spin connection $\om^{ab}$ has more components than the vielbein $e^a$, thus the Lorentz invariant first-order form contains redundant variables that must be eliminated before the constraint analysis can be performed.

This concretely demonstrates that the existence of ghost--free mass terms are not the obstruction to a charged spin-2 field, but instead the Einstein-Hilbert terms are incompatible with the requisite $U(1)$ structure needed to support a charged spin-2 theory.  Thus, having a ghost-free theory of a massive, self-interacting spin-2 field does not help one write down a theory of a ghost-free theory of a self-interacting, charged spin-2 field.

\acknowledgments
We would like to thank Shuang--Yong Zhou and Raquel Ribeiro for useful comments on the manuscript, and Kurt Hinterbichler for useful discussions. CdR is supported by Department of Energy grant DE-SC0009946. AJT are supported by a Department of Energy Early Career Award DE-SC0010600. AM is supported by the NSF Graduate Research Fellowship Program.
The authors would like to thank the Perimeter Institute for Theoretical Physics for hospitality and support during part of this work.

\appendix

\section*{Appendices}

\section{Three-dimensional Einstein-Cartan formalism}
\label{app:3D-Gravity}

The vielbein formalism has already been shown to greatly simplify the form of the interactions of ghost--free massive gravity and multi-gravity theories \cite{Nibbelink:2006sz,Chamseddine:2011mu,Hinterbichler:2012cn}.
Since we will be interested in modified kinetic terms in this work, we will be including the spin connection in our ADM analysis. This can be done using the Einstein-Cartan (EC) formalism, where the spin connection $\om^{ab}$ is treated as an independent field.

The EC formalism is particularly simple in three dimensions, which is why we focus on three dimensions.
\begin{itemize}
	\item[1.)] In a $D$-dimensional spacetime, the Hamiltonian analysis of the EC action is, in general, very complicated.  This is because the kinetic terms in the Hamiltonian go as $\dot{e}^a_i \om^{bc}_j \ep^{ij}\ep_{abc}$; therefore the spin connection $\om^{ab}$ is the momenta conjugate to $e^a$.  However, the number of spin connections $\om_i^{ab}$, which is $D(D-1)/2 \times (D-1)$, is in general much larger than the number of genuine conjugate momenta to $e^a_i$, which is $D \times (D-1)$.  To reconcile this, one will find that there are many secondary, second-class constraints that project out the excess of conjugate momenta and return the theory to the healthy number of phase space dofs.  Such an analysis is quite copious even for ordinary gravity \cite{Kiriushcheva:2009nj}.  Contrarily, it is uniquely true in $D=3$ that the conditions become just right and the number of spin connections exactly equals the number of conjugate momenta.  This makes the analysis of potentially ghostly interactions for gravity theories ideal in $D=3$.  The na\"ive expectation is that if the theories fail in $D=3$, a compactification argument tells us that they are unlikely to work in any higher dimensions (see the discussion in Sec.~\ref{sec:non-linear-completion}).

\item[2.)] In three dimensions, we are greatly aided by the \poin duality, which relates 1-forms, i.e. vectors, with 2-forms by the Hodge star, i.e. $\star (B_{\rho\si}) = \ \ep_{\mu}\,^{\rho\si}B_{\rho\si}$.  Using these tricks, we can define a dual spin connection $\om^a \equiv \ep^{abc} \om^{bc}$ which naturally comprises the conjugate momenta to $e^a$, rather than its more complicated form $\om^{ab}$.
\end{itemize}
This will cause the Hamiltonian analysis to simplify much more than in four or higher dimensions, however we emphasize that the main results of this paper are fully generalizable to arbitrary dimensions. 

\subsection{The EC Action in $D=3$}

To make this more concrete, let us start with the $D=3$ EC action\footnote{In what follows we will Wick rotate to Euclidean space, so the position of the indices does  matter.  Note that this is only true because we are working with the vielbein indices; if we were working with the spacetime indices, the difference is important.  One may trivially Wick rotate back to the Lorentzian by forcing upstairs indices to only contract with downstairs and interpreting it as the standard Minkowski inner product between them.} in differential form notation:
\ba
		S &=& M_3 \int  \ep_{abc} R^{ab} \wedge e^c \\
			&=& M_3  \int \ep_{abc} (\d\om^{ab} + \om^{ad} \wedge \om^{db}) \wedge e^c\,.
\ea
Next we define the dual of $\om^{ab}$ as $\om^a \equiv \ep^{abc} \om^{bc}$. The inverse is given by $\om^{ab} = \frac{1}{2} \ep^{abc}\om^c$
Then, distributing the overall $\ep_{abc}$ into the two terms and applying the definition of $\om^a$, one derives
\be
S = M_3 \int \left( \d\om^a - \frac{1}{4}\ep^{abc} \om^a \wed \om^b \right)\wed e^c\,,
\ee
This leads us to define the dual Riemann tensor
\be
		R[\om]^a = d\om^a - \frac{1}{4} \ep^{abc} \om^b \wedge \om^c\,.
\ee
Similarly, we can express the covariant derivative of a Lorentz vector $\la^a$ in terms of the dual spin connection as
\ba
	\mathcal{D} \la^a &\equiv& \d\la^a + \om^{ab}\la^b \, , \\
	&=& \d\la^a - \frac{1}{2} \ep^{abc} \om^b \la^c.
\ea
Then the EC action is
\be
S = M_3 \int R[\om]^a \wed e^a.
\ee
Varying this with respect to $e^a$ yields the Einstein equation
\be
R^a = 0.
\ee
Meanwhile varying this with respect to $\om$ gives the torsion free condition
\be
\mathcal{D} e^a = 0.
\ee

\subsection{Hamiltonian of EC Gravity in $D=3$}

We will now convert the EC action in the previous section into its Hamiltonian form.  After  integrating the exterior derivative by parts
\ba
		S  = M_3 \int\( \om^a \wedge \d e^a - \frac{1}{4}\ep^{abc} \om^a \wedge \om^b \wedge e^c\)\,.
\ea
In index notation this is given by
\be
		S = M_3 \int \d^3x\, \ep^{\mu\nu\rho} \left(\om^a_{\mu} \p_{\nu} e^a_{\rho} - \frac{1}{4} \ep_{abc} \om^a_{\mu} \om^b_{\nu} e^c_{\rho}\right).
\ee
We then perform the $(2+1)$-split onto the action, yielding
\ba
S		= M_3 \int \d^2x\ \d t\, \ep^{ij}\left[  \om^a_{i} \dot{e}^a_{j} + e^a_0 \left(\p_i\om^a_j -\frac{1}{4} \ep^{abc}\om^b_i \om^c_j \right) + \om^a_0 \left( \p_i e^a_j - \frac{1}{2}\ep^{abc} \om^b_{i} e^c_{j} \right) \right]. \qquad\quad
\ea
Here we see that $e^a_0$ and $\om^a_0$ enter into the theory as Lagrange multipliers, and given the definition of conjugate momenta
\be
\Pi^i_a = \frac{\p\mathcal{L}}{\p\dot{e}^a_i} = \ep^{ij}\om_j^a,
\ee
we see that the $\om^a$ are the momenta conjugate to $e^a$ as promised. The inverse Legendre transformation then easily shows us that the Hamiltonian is given by
\be
	 H[e,\om]  = M_3 \ep^{ij} \left[ e^a_0 \left(\p_i\om^a_j -\frac{1}{4} \ep^{abc}\om^b_i \om^c_j \right) + \om^a_0 \left( \p_i e^a_j - \frac{1}{2}\ep^{abc} \om^b_{i} e^c_{j} \right) \right],
\ee
which is pure constraint. This is expected because all diffeomorphism invariant theories give rise to Hamiltonians that are pure constraint.

\section{Application of first-order constraint analysis to bi-gravity}
\label{app:bi-gravity}
As a check on the method described in section~\ref{sec:non-minimal-extensions}, we will here show that the method can be used to show the absence of the Boulware-Deser mode in bi-gravity in three dimensions.
Start with bi-gravity with no cosmological constants
\be
S =M_3 \int \ep_{abc} \left( R[\om_1]^{ab} \wedge e_1 + R[\om_2]^{ab} \wedge e_2 + m^2 \left( c_1 e_1^a \wedge e_1^b \wedge e^c_2 + c_2 e_1^a \wedge e_2^b \wedge e_2^c \right)  \right) .
\ee

As in other sections, it is useful to work with the dual of the spin connection by defining $\om^a = \ep^{abc} \om^{bc}$.
Then we perturb to quadratic order around an arbitrary background
\ba
e^a_{I,\mu} &=& \bar{e}^a_{I,\mu} + v^a_{I,\mu} \nn \\
\om^{ab}_{I,\mu} &=& \bar{\om}^{ab}_{I,\mu} + \mu^{ab}_{I,\mu}.
\ea
Next we introduce the \stu fields at the level of the perturbations. We introduce the \stu fields through site 2 for convenience
\ba
v_2^a &\rightarrow& v_2^a + \bar{\mathcal{D}} \phi^a \nn \\
\mu_2^{ab} &\rightarrow& \mu_2^{ab} + \bar{\mathcal{D}} \lambda^{ab},
\ea
where $\bar{\mathcal{D}} = \d + \bar{\om}_1$ is the background covariant derivative. Since we have introduced the \stu fields that act as maps from site 2 to site 1, in this representation we may identify the diagonal local Lorentz transformations with site 1, and thus the spin connection appearing in $\bar{\mathcal{D}}$ is the spin connection for site 1.\\

The quadratic action takes the form (assuming the torsion vanishes, $\bar{\mathcal{D}}e=0$)
\ba
S^{(2)} &=& \int \ep_{abc}\left( \bar{\mathcal{D}} \mu_1^{ab} \wed v_1^c + \bar{\mathcal{D}} \mu_2^{ab} \wed v_2^c +\bar{\mathcal{D}} \lambda^{ab} \wed \bar{R}^{cd} \phi^d + 2 \bar{\mathcal{D}} \lambda^{ad} \wed \mu^{db}_2 \wed \bar{e}^c_2
\right)\nn \\
&& + m^2 \ep_{abc} \left[ c_1 \bar{\mathcal{D}} \phi^a \wed v_1^b \wed \bar{e}_1^c + c_2 \left( 2 \mathcal{\bar{D}} \phi^a \wed v_2^b \wed \bar{e}_1^c + 2 \bar{\mathcal{D}} \phi^a \wed v_1^b \wed \bar{e}_2^c
\right)\right]
+ S_{\rm N.D.} \nn 
\ea
where $S_{\rm N.D.}$ refers to terms with no derivatives on fluctuations.

We now define the duals $\mu^a \equiv \ep^{abc} \mu^{bc}$ and $\lambda^a \equiv \ep^{abc} \lambda^c$. Focusing on the time derivatives yields
\ba
S^{(2)} &=& \int \d^3 x\  \ep^{ij} \left( \dot{\mu}^a_{1,i} v^a_{1,j} + \dot{\mu}^a_{1,i} v^a_{1,j} + 2 \ep_{abc} \bar{e}_{2,j}^a \dot{\lambda}^b \mu_{2,i}^c  +\bar{R}^{ab}_{2,ij} \dot{\lambda}^a \phi^b \right) \nn \\
&& + m^2 \ep_{abc} \ep^{ij}\left[  \dot{\phi}^a v_{1,i}^b \left( c_1 \bar{e}^c_{1,j} + 2 c_2 \bar{e}^c_{2,j}\right) + 2 c_2  \dot{\phi}^a v_{2,i}^b  \bar{e}^c_{1,j}  \right]  + S_{N.D.}\nn \\
&=& \int \d^3 x\  \frac{1}{2} \xi_A \Om_{AB} \dot{\xi}_B - H.
\ea
In this case, $\Om_{AB}$ is a $30\times 30$ matrix. It is given by
\ba
\Om = \left(
\begin{array}{c | cccccc}
& \dot{\mu}^b_{1,j} & \dot{v}^b_{1,j} & \dot{\mu}^b_{2,j} & \dot{v}^b_{2,j} & \dot{\phi}^b & \dot{\lambda}^b \\
\hline
\mu^a_{1,i} 	&  & - \ep_{ij} \delta^{ab} &  &  &  &  \\
v^a_{1,i} 		& \ \ep_{ij} \delta^{ab}\  & & & & m^2 \mathcal{A}^{ab}_i\left[c_1 \bar{e}_1+ 2 c_2 \bar{e}_2\right] & \\
\mu^a_{2,i} 	& & & & - \ep_{ij} \delta^{ab}  & & -\mathcal{A}^{ab}_i[\bar{e}_2] \\
v^a_{2,i} 		& & &\ \ep_{ij} \delta^{ab} \  & & 2 m^2 c_2 \mathcal{A}^{ab}_i[\bar{e}_1] & \\
\phi^a 		&  &-m^2 \mathcal{A}^{ab}_j\left[c_1 \bar{e}_1 + 2 c_2 \bar{e}_2\right]  & & -2 m^2 c_2 \mathcal{A}^{ab}_j[\bar{e}_1] & &  \ep^{ij}\bar{R}^{ab}_{2,ij} \\
\lambda^a	&  &  & \ \mathcal{A}^{ab}_j[\bar{e}_2] &  &  - \ep^{ij} \bar{R}^{ab}_{2,ij} &
\end{array}
\right) \nn
\ea
where the background-dependent function $\mathcal{A}^{ab}_i[e]$ is given by
\be
\mathcal{A}^{ab}_i[e] =  \ep_{abc} \ep_{ij} \bar{e}^c_j.
\ee
Computing the eigenvalues of $\Om_{AB}$ explicitly, we find that there are 2 eigenvalues that vanish identically, independently of the choice of background and of the parameter choices. This is a proof, in 3 dimensions, that ghost-free bi-gravity (and thus ghost-free massive gravity) propagates no more than two dofs around any arbitrary off shell background. This method is very simple.

Of course there are background for which there are more than 2 zero eigenvalues. This corresponds to the well known backgrounds in the literature where the kinetic term for one or more of the perturbations vanishes, signaling a strongly coupled background solution.

In \cite{Banados:2013fda}, the analysis was done in unitary gauge and it was pointed out that there is an ambiguity corresponding to the need to impose a secondary constraint. That ambiguity corresponds here to the way we introduce the \stu fields. After introducing the Lorentz \stu fields, we may always chose a gauge where the symmetric vielbein condition $e^a_{[\mu} f^a_{\nu]} = 0$ is satisfied.

\section{Alternative approach to Hamiltonian analysis}
\label{app:flat-space-argument}
In this appendix we provide an alternative, perhaps more direct argument that the new kinetic terms that we were forced to introduce by the $U(1)$ symmetry reintroduce the Boulware-Deser ghosts. The outline of the argument is
\begin{itemize}
\item We will start with the $U(1)$ invariant action suggested by deconstruction. We will introduce 2 copies of the Lorentz and diff \stu fields so that we reintroduce the full undiagonalized gauge symmetries. As a result, all constraints will be first class.

\item By the counting argument of section~\ref{sec:counting}, we will see that the theory will only propagate 4 dofs (the correct number for two massive gravitons in 3 dimensions) if one linear combination of the \stu fields on each site is non-dynamical.

\item By perturbing the action to cubic order about Minkowski space, we will see that for a generic choice of parameters that all of the \stu fields will be dynamical, and so the theory will propagate too many dofs. We may identify these extra propagating modes as Boulware-Deser ghosts.
\end{itemize}

A key feature of our analysis is that all of the constraints are first class. As a result, we will not generate any new secondary constraints.

\subsection{Perturbation theory in the \stu language}
Our starting point is the Deconstruction-inspired theory written in site language, given by Equations~(\ref{eq:u1-mass-term-site-language}--\ref{eq:u1-kinetic-term-site-language}). We showed that this was equivalent to the non-linear ansatz in Equation~\eqref{eq:wedgy-action-3}.
We first introduce the \stu fields for both diff and local Lorentz symmetries \cite{Ondo:2013wka}
\ba
&& e^a_{I,\mu}(x)  \rightarrow  \partial_\mu \Phi^{\mu'}_{I} \Lambda^{a a'}_I e^{a'}_{\mu'}(x), \ \ \ \ \ \ \ \ \ \ \ \ \ \ \ \ \ \ \ \ \ \ \ \ \  I=2,3  \\
&& \omega^{ab}_{I,\mu}  \rightarrow  \partial_\mu \Phi^{\mu'}_I \left(\Lambda^{a a'} \omega^{a' b'}_{I,\mu'} \Lambda^{b' b}_I - \Lambda^{a c}_{I} \partial_{\mu'} \Lambda^{c b}_I\right),  \ \ I=2,3
\ea
Note that we only introduce \stu fields on sites 2 and 3. In this way we associate the diagonal copies of the gauge symmetries with the gauge symmetries acting on site 1. The choice of site 1 here is arbitrary but convenient for the analysis.

The $\Phi^\mu_I$ are maps from site $I=2,3$ to site 1, which has the coordinates $x^\mu$,
\be
Y_I^\mu(x_I^\mu) = x^\mu.
\ee
Thus diff indices are raised and lowered with the metric on site 1. We will now work perturbatively around flat space
\ba
&& e^a_{I,\mu} = \delta^a_\mu + h^a_{I,\mu}  \, \nn \\
&& \om^{ab}_{I,\mu} = \theta^{ab}_{I,\mu} = \frac{1}{2} \ep^{abc}\theta^c_I  \, \nn \\
&& \Phi^\mu_I = x^\mu + B^\mu_I  \, \nn \\
&& \Lambda^{ab}_I = e^{\lambda^{ab}} = \delta^{ab} + \lambda^{ab} + \cdots = \delta^{ab} + \frac{1}{2} \ep^{abc} \lambda^c.
\ea
Note that $\lambda^{ab}_I = -\lambda^{ba}_I$. Also note that we have used the fact that we are in three dimensions to rewrite antisymmetric tensors with 2 indices as vectors, $V^a = \ep^{abc} V^{bc}$.
We will now work in units where $\Mpl=1$, and work perturbatively in the variables above. 

\subsection{Determining the size of the na\"ive phase space}
\label{app:symplectic-form}
To implement the counting of section~\ref{sec:counting}, it is thus necessary to establish the size of the na\"ive phase space. In other words, we must count the number of dynamical variables before any constraints are imposed. The form structure of the action guarantees that, to cubic order, the action takes the form
\be
\label{eq:first-order-action}
S^{(2)} + S^{(3)} = \int \d^3x \( \xi_m \Om[\xi]_{m n} \dot{\xi}_n - H(\xi)\)\,,
\ee
where the $\xi_m$ are the dynamical variables. When we introduce our new kinetic terms, we will find that the phase space measure $\Om[\xi]$ is not written in Darboux form. In other words, it will not be possible to cleanly separate the fields into coordinates and conjugate momenta without doing a field redefinition.

To avoid needing to explicitly find the field redefinition to go to Darboux form (which is always possible locally), we will determine the na\"ive phase space directly from the symplectic form $\Om$. By varying the action with respect to $\xi_m$, we obtain the equations of motion
\be
\Om_{mn} \dot{\xi}^n = \frac{\partial H}{\partial \xi^m}.
\ee
This is a set of dynamical equations. If $\Om$ is invertible, then all of the $\xi_n$ have independent, dynamical equations. If $\Om$ is not invertible, then not all of the equations are independent. The number of nonzero eigenvalues of $\Om[\xi]_{mn}$ gives the number of dynamical variables in the na\"ive phase space. For more details see for example \cite{Faddeev:1988qp}.

\subsection{Counting degrees of freedom at quadratic order}
We now obtain the derivative parts of the action at quadratic order. We find
\be
S^{(2)} = S^{(2)}_{\rm GR} + S^{(2)}_m + S^{(2)}_\gamma\,,
\ee
where
\ba
S^{(2)}_{GR} &=& \int \d^3 x \sum_I  \partial_\mu \theta_{I,\nu}^a h_{I,\rho}^a + S^{(2)}_{\rm GR,N.D.},  \\
S^{(2)}_{m} &=& \int \d^3 x \sum_{IJK} 3! m^2 \beta_{IJK} \Big( \pa_a B^a_I [h_J] - \frac{1}{2} \ep^{abc} \pa_a B_{I,b} \lambda_{J,c} - \pa_a B_{I,b} h_J^{ab}  \\
&& + \frac{1}{2} \ep^{abc} \pa_a B_{I,b} \lambda_{I,c} + \pa_a B_{I,b} h^{ab}_I \Big) + S^{(2)}_{m, {\rm N.D.}}, \nn \\
S^{(2)}_\gamma &=& 2 \int \d^3 x \sum_{IJK} \lambda^a_I \left(\partial_a \theta^b_{J,b} - \partial_b \theta^b_{J,a}\right) + S_{\gamma, {\rm N.D.}}^{(2)},
\ea
where the subscript ${\rm N.D.}$ indicates terms with no derivatives that are irrelevant for this analysis.

The main thing to do is to count the number of dofs. We note that because of the form structure, at quadratic order, the action is already in first order form
\be
S^{(2)} = \int \d t \sum_n p_n \dot{q}_n - H,
\ee
Using the tadpole cancellation condition
\be
\sum_{JK} \beta_{IJK} = 0.
\ee
as well as
\be
\sum_K \gamma_{IJK} = 0,
\ee
this becomes
\ba
S^{(2)} &=& \int \d^3 x \sum_I \dot{\theta}^0_{I,i} \left( \ep^{ij} h^0_{I,j}  \right)  + \dot{\theta}^i_{I,j} \left(\ep^{jk} h^i_{I,k} \right) \nn \\
&&+ \dot{B}_I^0 \left(6m^2 \sum_{JK} \beta_{IJK}\left(h^i_{J,i} \right)\right) + \dot{B}_I^i \left(- 6m^2 \sum_{JK} \beta_{IJK} \left( \eta_{ij} \ep^{jk}  \lambda_J^k  + h^0_{J,i} \right) \right) \nn \\
&& - H.\nn
\ea
We see that at quadratic order, $B^0_I$ does not have an independent conjugate momentum.
\ba
\left(\pi_\theta \right)^\mu_{I,a} &=&  \ep^{0\mu\nu} h_{I,a \nu}  \\
\left(\pi_B\right)^\mu_{I} &=& 6 m^2 \sum_{JK} \beta_{IJK} \left(\delta^\mu_0 [h_J] - \frac{1}{2} \ep^{0\mu\nu} \lambda_{J\nu} - h^{0\mu}_J + \frac{1}{2} \ep^{0\mu\nu} \lambda_{I,\nu} \right).
\ea
In particular, notice that
\be
(\pi_\theta)^0_{I,a} = 0,
\ee
reflecting the fact that $\theta_{I,0}^a$ is the Lagrange multiplier for a set of first class constraints.
Also notice that
\be
(\pi_B)^0_I = 3! m^2 \sum_{JK} \beta_{IJK} h^i_{J,i} = \frac{3!}{2} m^2 \sum_{JK} \beta_{IJK} \epsilon^{0ai} \left(\pi_\theta\right)_{ai}.
\ee
This is a crucial step. $B_0$ does not have an independent conjugate momentum. If it did, $B_0$ and its conjugate momentum would be the dynamical variables representing the Boulware-Deser ghost.

It is useful to diagonalize the kinetic term to remove the momentum conjugate to $B^0_I$, by definining the diagonalized field $\chi$
\be
\chi^a_{I,i} = \theta^a_{I,i} - \frac{3!}{2} m^2 \sum_{JK} \beta_{IJK} \epsilon^{0ai} B_J^0.
\ee
The momentum conjugate to $\chi$ is just $\pi_\chi = \pi_\theta$.
Then the action takes the form
\be
S^{(2)} = \int \d^3 x \ \ \dot{\chi}^a_{I,i} \left(\pi_\chi \right)^i_{I,a} + \dot{B}^i_I \left(\pi_B\right)_{I,i} - H.
\ee
In this form, we see that $\chi^a_{I,0}$ and $h^a_{I,0}$ are Lagrange multipliers for first class constraints. We also see that $B^0,I$ and $\lambda^0_I$ are not dynamical, and they do not enter with any conjugate momentum.

This allows us to recover the tri-gravity result, where we should not have any Boulware-Deser modes present in the theory.
Note that the kinetic term generated by applying the $U(1)$ projection to the action from deconstruction falls into this category. That was indeed crucial to reproduce the Federbrush action. This is a reflection of the fact that the charge violating operators only arise at cubic order.

\subsection{Counting degrees of freedom at cubic order}
At cubic order the action is
\be
S^{(3)} = S^{(3)}_{\rm GR} + S^{(3)}_m + S^{(3)}_\gamma.
\ee
$S^{(3)}_{\rm GR}$ has no time derivatives because of the form structure. Meanwhile,
\ba
S^{(3)}_{m} &=& 3m^2 \int \d^3 x \sum_{IJK} \beta_{IJK} \Big( \ep_{abc} \ep^{\mu\nu\rho} \left[\pa_\mu B_I^a \pa_\nu B^b_J \left(h^c_{K,\rho} + \lambda^c_{K,\rho} \right)\right] + \pa_\mu B_I^a \left(\lambda_{J,\nu}^b + h_{J,\nu}^b \right)\left(\lambda^c_{K,\rho} + h^c_{K,\rho}\right) \nn \\
&&+ 2 \left[\pa_a B_{I,a'} \pa_b B_J^b \left(\lambda^{aa'}_I + h^{aa'}_I \right) - \pa_a B_{I,a'} \pa_b B_J^a \left(\lambda^{ba'}_I + h^{ba'}_I \right) \right] \nn \\
&& - 2\pa_b B_{I,a'} \lambda_I^{aa'} \lambda_J^{ba} + \pa_a B^a_I \lambda^{bq}_J \lambda^q_{J,b} - \pa_b B_{I,a} \lambda_J^{bq} \lambda_J^{qa} + 2\pa_a B_{I,a'} h^{aa'}_I [h_J] \nn \\
&& - 2 \pa_b B_{I,a'} h^{aa'}_I h^{ba}_J - 2 \pa_b B_{I,a'} h_I^{aa'} h_J^{ba} + 2 \pa_a B_{I,a'} \lambda^{aa'}_I [h_J] - 2 \pa_b B_{I,a'} \lambda^{aa'}_I h_J^{ba}\Big) \nn \\
&\sim& m^2 \int \d^3 x \left(\partial B\right)^2 (h+\lambda) + \partial B (\lambda^2 + \lambda h + h^2).
\ea
Finally, using the condition that $\sum_{IJK}\gamma_{IJK}=0$,
\ba
S_\gamma^{(3)} &=& \int \d^3 x \sum_{IJK} \gamma_{IJK} \ep_{abc} \ep^{\mu\nu\rho} \Bigg( \pa_\mu \la_I^{ap} \left(\pa_\nu \la_J^{pb} - 2 \theta_{J,\nu}^{pb} \right)\left(\la_{K,\rho}^c + h_{K,\rho}^c \right) \nn \\
&& + \left(\theta_{I,\mu}^{ap} - 2 \pa_\mu \la_I^{ap} \right) \theta_{J,\nu}^{pb} \pa_\rho B^c_K \Bigg) \nn \\
&\sim& \int \d^3 x \gamma \left[ (\partial \lambda)^2(\lambda + h) + \partial \lambda \theta (\lambda + h + \partial B) + \theta^2 \partial B \right].
\ea
We see that to this order, the action is still in first order form, with one time derivative per field. However, $B,\lambda$, and $\omega$ all appear with time derivatives, and it is not possible to integrate by parts so that only two of them contain time derivatives. Thus the introduction of our new kinetic interaction has taken the action out of Darboux form, and instead the action is written in a more general form. Thus to determine the size of the na\"ive phase space, we must determine the number of nonzero eigenvalues of the phase space measure $\Omega$.

\subsubsection{Form of $\Om$}
The crucial question is whether or not $\Om_{mn}$ is invertible. If it is, then the na\"ive phase space contains all of the fields as potential dofs. If it is not, then some of the fields are not dofs. We have seen that we will not propagate the correct number of dofs for a massive spin-2 field unless some of the freedom is projected out.

As stressed above, the Boulware Deser mode is associated with some of the components of the \stu fields. Thus in order to remove the Boulware-Deser ghost, it is crucial that $\det \Om = 0$.
Since we are working perturbatively, we may write
\be
\Om = \Om^{0} + \ep \Om^{(1)} + \cdots\,,
\ee
where the superscript indicates the order in the field. The constant part $\Om^{(0)}$ is determined from the quadratic action, the part linear in the fields is determined from the cubic action.

The form of $\Om$ is
\be
\Om = \left(
\begin{array}{ccc|c}
 &  									&  			&  				\nn \\
 & \Om^{(0)}_{ij} + \ep \Om^{(1)}_{ij} 		& 			& \ep\Om^{(1)}_{ai}	 \nn \\
 &  									&  			& 			 	\nn \\
 \hline
 & - \ep \Om^{(1)}_{ai} 					&  			& \ep \Om^{(1)}_{ab}
\end{array}
 \right)\,,
\ee
where $i,j$ run over $h^a_{I,i}, \chi^{a}_{I,i}$, and $a,b$ run over $\lambda_0^I, B_0^I$.

The determinant of this matrix can be computed perturbatively as
\be
\det \Om = \det\left(\Om_{ij}^{(0)}\right) \times \det\left( \Om_{ab}^{(1)}\right) \ep^4 + \mathcal{O}(\ep^5).
\ee
It is not necessary to compute $\det \Om^{(0)}_{ij}$ explicitly, it is enough to know that it is nonzero. The reason it is nonzero is because at quadratic order, all of the fields that the $i,j$ indices run over are dynamical.

Now $\Om^{(1)}_{ab}$ in principle will have contributions from the mass and kinetic terms
\be
\Om^{(1)}_{ab} = \Om^{(1),m}_{ab} + \Om^{(1),\gamma}_{ab}.
\ee
However an explicit calculation shows that
\be
\Om^{(1),m}_{ab} = 0,
\ee
consistent with the expectation that $\det \Om=0$ for ghost-free tri--gravity.

\subsubsection{Computing $\det \Om^{(1)}_{ij}$}

Since $\det \Om=0$, we only need to compute $\Om^{(1),\gamma}_{ab}$. This is a linear function of all of fields $\lambda^0_I, B^0_I, \chi^a_{I,i}$
\be
\Om^{(1)}_{ab} = \frac{\delta \pi_a}{\delta \xi^b} = \sum_I \left(A_{\lambda,I}\right)_{ab} \lambda^0_I  + \left(A_{B,I}\right)_{ab} B^0_I + \left(A_{\chi,I}\right)_{ab}^{ia} \chi^a_{I,i},
\ee
where $\xi^a = \{\lambda^0_I, B^0_I\}$ and where $A_\lambda,A_B, A_\chi$ are $4\times 4$ matrices of field independent coefficients.

Because in a healthy theory the Boulware-Deser ghost must be absent from all solutions, we need only find one solution for which $\Om$ is invertible. Thus we will consider the case that $B^0_I = \chi^a_{I,i} = 0$. We emphasize that this may only be done after computing $\Om$.

Now the schematic forms of the momenta are
\ba
\pi_{B^0} &\sim & \partial B (h + \lambda) + \lambda^2 + \lambda h + h^2 \,  \nn \\
\pi_{\lambda^0} &\sim& B \partial B + B \lambda + B h.
\ea
Then $A_\lambda$ takes the form
\be
A_\lambda \sim \left(
\begin{array}{cc}
\frac{\pa \pi_{B^0}}{\partial B^0} - \frac{\pa \pi_{B^0} }{\pa \pi_{B^0} }& \frac{\pa \pi_{B^0}}{\partial \lambda^0} - \frac{\pa \pi_{\lambda^0} }{\pa \pi_{B^0} } \\
\frac{\pa \pi_{\lambda^0} }{\pa \pi_{B^0} } - \frac{\pa \pi_{B^0}}{\partial \lambda^0} & \frac{\pa \pi_{\lambda^0}}{\partial \lambda^0} - \frac{\pa \pi_{\lambda^0} }{\pa \pi_{\lambda^0} }
\end{array}
\right)\Bigg|_{B^0_I = \chi^a_{I,i} = 0}.
\ee
Then we find that, when $B^0_I = \chi^a_{I,i} = 0$, the $4 \times 4$ matrix $\Om_{ab}^{(1)}$ can be written in terms of the simpler $2\times 2$ matrix $\mu$ as
\be
\Om^{(1)}_{ab} |_{B,\chi=0} =\left(
\begin{array}{cc}
0 & \mu \nn \\
-\mu & 0
\end{array} 
\right) \, ,
\ee
where
\be
\mu_{IJ} =  \frac{2 \times 3!}{(2c)^4} m^2  \sum_{PQR} \gamma_{IPQ} \beta_{PJR} \lambda^0_Q, \ \ \ I,J = 1,2 \, ,
\ee
then
\be
\det \Om^{(1)}_{ab} = \det(\mu)^2.
\ee

Since $\gamma_{IJK}$ and $\beta_{IJK}$ are fixed, we find
\be
\det(\mu) = -\frac{363}{100} \left(\lambda_{0,2} - \lambda_{0,3}\right)^2.
\ee
There are clearly solutions for which this is non-zero, signaling the presence of a Boulware-Deser mode.

\bibliographystyle{JHEPmodplain}
\bibliography{refs_2}

\end{document}